\documentclass[letterpaper,journal]{IEEEtran}

\usepackage{mathtools}
\usepackage{amsmath} 
\allowdisplaybreaks

\usepackage{amsmath,amsfonts,amssymb}
\usepackage{algorithmic}
\usepackage{algorithm}
\usepackage{array}
\usepackage[caption=false,font=normalsize,labelfont=bf,textfont=rm]{subfig}
\usepackage{textcomp}
\usepackage{stfloats}
\usepackage{bbm}
\usepackage{url}

\usepackage{verbatim}
\usepackage{graphicx}
\usepackage{xcolor}
\usepackage{cite}
\usepackage{enumitem}
\usepackage{xurl}
\usepackage{adjustbox}
\usepackage{threeparttable}
\usepackage{tabularx,booktabs,array,multirow,siunitx}
\newcolumntype{L}[1]{>{\raggedright\arraybackslash}p{#1}}
\newcolumntype{Y}{>{\raggedright\arraybackslash}X}
\usepackage[colorlinks=true,allcolors=blue]{hyperref}
\hypersetup{breaklinks=true}
\begin{document}

\title{Large-Scale Data Center Integration in Power Systems: Uncertainty and Multi-Scale Variation}

\author{Le~Xie,~\IEEEmembership{Fellow,~IEEE}, Dongjoo~Kim,~\IEEEmembership{Student Member,~IEEE}, Qian~Zhang,~\IEEEmembership{Student Member,~IEEE}, Grant~Ruan,~\IEEEmembership{Member,~IEEE}, and Paul~L.~Joskow

\thanks{\copyright~2026 IEEE. Personal use of this material is permitted. Permission from IEEE must be obtained for all other uses, in any current or future media, including reprinting/republishing this material for advertising or promotional purposes, creating new collective works, for resale or redistribution to servers or lists, or reuse of any copyrighted component of this work in other works.}
\thanks{Le~Xie, Qian~Zhang are with the Harvard John A. Paulson School of Engineering and Applied Sciences, Harvard University, Allston, MA, USA.}
\thanks{Dongjoo~Kim is with the Department of Electrical and Computer Engineering, Texas A\&M University, College Station, TX, USA.}
\thanks{Grant~Ruan is with the Laboratory for Information \& Decision Systems (LIDS), Massachusetts Institute of Technology, Cambridge, MA, USA.}
\thanks{Paul~L.~Joskow is with the Department of Economics, Massachusetts Institute of Technology, Cambridge, MA, USA.}
\thanks{Corresponding author: Le Xie, (e-mail: xie@seas.harvard.edu).}
}

\markboth{Accepted for publication in Proceedings of the IEEE.}%
{Xie \MakeLowercase{\textit{et al.}}: Large-Scale Data Center Integration in Power Systems}

\maketitle

\begin{abstract}
Hyperscale artificial intelligence (AI) data centers are emerging as a structurally new class of electricity demand. Unlike conventional loads, they materialize as spatially concentrated, transmission-level point loads with rapid deployment timelines and converter-dominated dynamics. Their integration is governed by two defining properties: forecast uncertainty in load realization and multi-scale temporal variation spanning sub-seconds to planning horizons.

We develop a unified framework that links reliability and affordability to these dimensions through a four-quadrant structure: (reliability, affordability) $\times$ (uncertainty, variation), which integrates probabilistic resource adequacy, generation and transmission expansion, operational security, and market design.

Measured traces from a public generative-AI workload dataset show strongly \emph{tonal} GPU power spectra rather than smooth broadband behavior: harmonic combs at the training-step cadence, and a distinct sub-second periodicity in inference. Across these traces the dominant variance band shifts with workload type---seconds for training and offline inference, tens of seconds and longer for online inference. We report this as a characterization of the measured workloads rather than a universal signature, and take it to indicate that such loads warrant a spectral description rather than a peak-demand one. These signatures directly affect reserve requirements, ramping products, voltage stability, ride-through performance, and harmonic interactions in converter-rich grids.

Taken together, the analytical framework and empirical results establish a system-level basis for assessing the reliability and cost implications of large AI data center integration across planning and operational horizons.
\end{abstract}

\section{Introduction}\label{sec:introduction}
Electricity demand from computing is projected to grow several-fold over the next decade, driven by the rapid expansion of artificial intelligence (AI) training and inference workloads and periodic surges and geographic shifts in cryptocurrency mining. However, these forecasts remain uncertain and reflect variability in AI adoption trajectories, improvements in hardware efficiency, and the cyclical nature of cryptocurrency markets. This growth represents a fundamental shift in electricity consumption, but it is not uniform across the data-center fleet. In practice, there are thousands of data centers, and the majority are small-to-medium facilities that typically interconnect at distribution grids and exhibit load profiles similar to other commercial customers. The bulk power system planning and operation challenges, however, are driven disproportionately by hyperscale data centers: spatially lumpy, megawatt-to-gigawatt-class campuses that often pursue transmission-level (or sub-transmission) interconnection and develop substantial ``inside-the-fence'' infrastructure, including on-site distribution, short-duration storage/uninterruptible power supply (UPS) systems, and backup generation. The magnitude and pace of this transition are already evident: major forecasting organizations project U.S. data-center electricity consumption to reach hundreds of terawatt-hours by 2030, while individual hyperscale AI facilities can approach gigawatt-scale demand, comparable to the capacity of a large nuclear unit or the peak load of a mid-sized city.

This transition from cryptocurrency mining to AI-driven computing has amplified both the scale and the complexity of grid integration challenges. While crypto mining facilities demonstrated that computing workloads could materialize rapidly as large point loads with minimal geographic diversification, AI data centers introduce additional dimensions of uncertainty and variation that differ in terms of their economic drivers, operational flexibility, and long-horizon predictability. Compared with the rule-based predictability of crypto mining, AI demand growth is driven by factors largely exogenous to traditional load forecasting models~\cite{ZhangChungXie2026DataCenterTransition}.

The rapid construction of large-scale AI data centers is transforming electricity demand in ways that challenge conventional power system planning and operation. These facilities can arrive as concentrated, high-power point loads on short timelines (often within 12--24 months), while generation and transmission additions typically require longer lead times of 3--7 years. This temporal mismatch creates new stress for both grid reliability and customer affordability. Moreover, AI data centers exhibit pronounced spatial concentration, which amplifies the localized impact of forecast errors and operational variability. These spatial features, combined with the temporal uncertainty of project duration and the multi-scale variation of operational profiles, create a fundamentally different risk landscape from that of traditional industrial/commercial loads.

In this paper, we interpret reliability broadly as encompassing three interrelated dimensions: i)~resource adequacy: planning reserve and energy adequacy, ii)~operational security: deliverability and contingency-secure operation, and iii)~stability: voltage and frequency stability, and oscillations. We argue that two properties dominate the landscape of AI data center integration: uncertainty (forecast ambiguity in size, timing, location, and firmness) and variation (multi-scale temporal dynamics from sub-second to minutes and beyond). These properties manifest across different time horizons and interact with both reliability and affordability concerns, which motivate a structured analytical framework that can systematically address the challenges and opportunities of large-scale data center integration.

Recent studies have explored the grid integration issues of data centers, specifically focusing on demand response potential and ancillary service provision~\cite{Wang2019}, power quality and harmonic distortion~\cite{ieee_519_2022}, and planning methods for accommodating large loads~\cite{epri_large_load_2024}. However, existing efforts address only one of the uncertainty and variation aspects, without explicitly connecting how these properties interact across time scales and propagate into both reliability and affordability outcomes. Moreover, most works treat data centers as homogeneous entities, without distinguishing how AI training workloads (synchronized, step-like dynamics) differ fundamentally from inference workloads driven by bursty user requests, or how these operational signatures translate into practical metrics and planning requirements.

The separation between engineering and economic studies is another notable gap. Engineering studies of dynamic performance, power quality, and stability often do not address cost allocation or tariff design questions, while economic and policy analyses of cost recovery and rate design seldom capture the load variability that determines infrastructure upgrades. This disciplinary division can lead to suboptimal outcomes: tariffs that fail to align incentives with operational needs, or infrastructure investments that deviate from the least-cost portfolio. There are clear concerns regarding the treatment of uncertainty: While stochastic optimization is popular in generation expansion planning, its application to demand-side uncertainty remains largely limited and cannot capture the discrete, spatially concentrated demand with uncertain timing and scale.

This paper bridges these gaps through three main contributions. First, we develop a unified analytical framework that explicitly connects uncertainty (forecasting challenges in size, timing, location, and firmness) and variation (multi-scale temporal dynamics from sub-seconds to years) to both reliability and affordability outcomes. This framework is structured as a four-quadrant matrix (see Table~\ref{tab:2x2_logic}): (reliability, affordability)$\times$(uncertainty, variation), which systematically organizes the existing challenges, evidence, and mitigation strategies across these dimensions.

Second, we provide an empirical characterization of AI data center variability across multiple time scales, including spectral analysis of measured workload traces that reveals strongly \emph{tonal} signatures---harmonic combs at the training-step cadence and a distinct inference periodicity---and a dominant variance band that shifts with workload type. We present these as properties of the measured traces rather than as universal characteristics of AI workloads. This finding benefits both forecasting methodologies and reserve sizing. We connect these temporal signatures to practical metrics and operational requirements, establishing quantitative links between workload characteristics (training versus inference) and power-system performance.

Third, we synthesize evidence from multiple domains (planning models, operational experiences, tariff design practices, and dynamic performance requirements) to identify actionable techno-economic policy guidance. Rather than isolating the engineering, economic, and policy discussions, we explicitly examine how tariff and contracting choices affect which infrastructure portfolios are built, how uncertainty in load forecasts propagates into adequacy risks and cost allocation decisions, and how operational flexibility mechanisms can be designed to align incentives with system needs.

The contribution of this paper is integrative rather than technology-specific. Detailed dynamic simulation is indispensable for evaluating a particular data-center topology, control strategy, and grid condition, and a growing body of work---including several of our own prior studies---already addresses those questions in depth: load characterization and grid-impact reviews~\cite{Chen2025}, electromagnetic-transient models for ride-through assessment~\cite{Samanta2024}, large-signal data-driven impedance models~\cite{ruan2025data}, Lyapunov-certified adaptive dynamics models~\cite{jena2026lilad}, and ISO-scale flexibility-aware siting studies~\cite{kim2026siting}. A single additional case study would represent one system under one operating condition and could not validate a framework that spans microsecond dynamics to multi-decade planning; we therefore ground the framework in this quantitative literature and in measured workload data.

The rest of this paper is organized as follows:

Section~\ref{sec:dc_newclass} establishes the foundational concepts of uncertainty and multi-scale variation by characterizing how AI data centers differ from conventional loads, introducing spatial-concentration metrics that contextualize why these issues matter for grid planning, and validating the multi-scale structure against measured generative-AI workload traces. Section~\ref{sec:adequacy_uncertainty} addresses reliability under uncertainty by examining how forecasting challenges propagate into resource adequacy risks, with separate treatment of systems with and without capacity markets. Section~\ref{sec:tariff_uncertainty} examines affordability under uncertainty, reviewing how tariff design and cost allocation mechanisms can discipline speculative load and limit cross-subsidies. Section~\ref{sec:security_variation} addresses operational security and stability under variation, covering both sub-second dynamic performance and seconds-to-minutes ramping requirements. Section~\ref{sec:affordability_variation} examines the incremental infrastructure and operational capabilities needed to accommodate fast variation and connects these engineering requirements to cost assessment and allocation. Section~\ref{sec:framework} outlines a techno-economic policy framework that integrates these dimensions and proposes directions for aligned incentives, commitments, and physical constraints. Finally, Section~\ref{sec:conclusion} concludes this paper with a concise summary.

\begin{figure*}[!t]
\centering
  \includegraphics[width=0.7\textwidth]{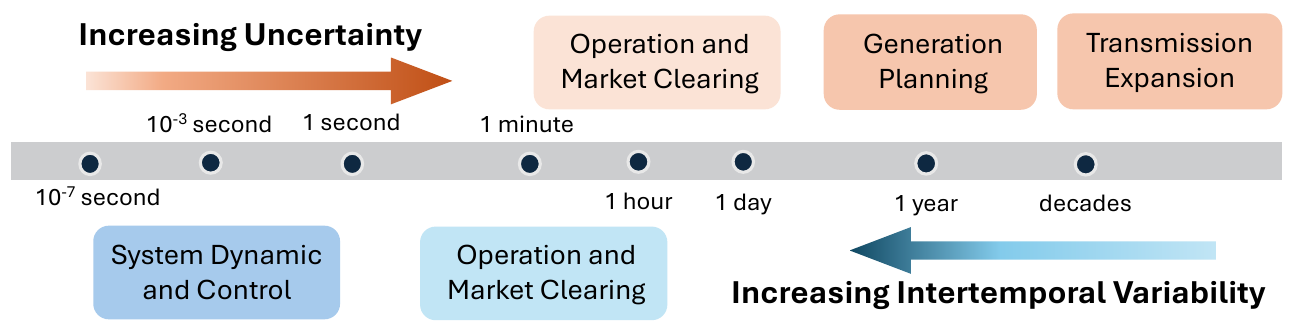}
  \caption{Timescale map linking the two defining properties of AI data-center demand to power-system functions. The upper orange progression denotes increasing forecast uncertainty as the horizon extends from operations and market clearing to generation planning and transmission expansion. The lower blue progression denotes increasing inter-temporal variation toward the shorter horizons governed by system dynamics and control and by operations and market clearing. Together, the two progressions motivate the reliability--affordability framework summarized in Table~\ref{tab:2x2_logic}.}
  \label{sec2overview}
\end{figure*}

\begin{table*}
\centering
\caption{The proposed four-quadrant logic: (reliability, affordability)$\times$(uncertainty, variation)}
\label{tab:2x2_logic}
\begin{tabular}{p{0.17\linewidth} p{0.37\linewidth} p{0.37\linewidth}}
\hline
\textbf{} & \textbf{Uncertainty} & \textbf{Variation} \\
\hline
\textbf{Reliability} (Resource Adequacy, Operational Security, Dynamic Stability) &
\begin{minipage}[t]{\linewidth}\vspace{2pt}
\textbf{Resource adequacy risk}: planning reserve errors, procurement mismatch;\\
\textbf{Deliverability/siting risk}: clustered load vs.\ network lead times;\\
\textbf{Mitigation}: milestone-based interconnection, probabilistic scenarios.\\
\emph{See Sec.~\ref{sec:adequacy_uncertainty}}
\vspace{2pt}\end{minipage}
&
\begin{minipage}[t]{\linewidth}\vspace{2pt}
\textbf{Operational security}: ramps, reserves, congestion;\\
\textbf{Stability}: ride-through, oscillations, protection interactions;\\
\textbf{Mitigation}: demand response, UPS/storage participation, dynamic modeling.\\
\emph{See Sec.~\ref{sec:security_variation}}
\vspace{2pt}\end{minipage}
\\
\hline
\textbf{Affordability} (New Facility Requirements, Cost Allocation) &
\begin{minipage}[t]{\linewidth}\vspace{2pt}
\textbf{Cost allocation under load risk}: stranded upgrades, bill volatility;\\
\textbf{Tariff design}: minimum bills, deposits, large-load classes;\\
\textbf{Equity}: limit unjustified cost shifts to other customers.\\
\emph{See Sec.~\ref{sec:tariff_uncertainty}}
\vspace{2pt}\end{minipage}
&
\begin{minipage}[t]{\linewidth}\vspace{2pt}
\textbf{Incremental infrastructure needs}: storage, reactive power (VAR) support, filtering, monitoring;\\
\textbf{System vs.\ behind-the-meter choice};\\
\textbf{New ramp products market}.\\
\emph{See Sec.~\ref{sec:affordability_variation}}
\vspace{2pt}\end{minipage}
\\
\hline
\end{tabular}
\end{table*}

\section{AI Data Centers as a New Class of Demand: Uncertainty and Multi-Scale Variation}\label{sec:dc_newclass}

\subsection{A New Class of Demand}

Conventional power system planning and rate design largely evolved under gradual growth of industrial customers whose operating schedules and energy use are comparatively stable. AI data centers depart from this paradigm in three interrelated aspects: i)~they materialize rapidly as large step changes at a small number of substations; ii)~the magnitude, timing, and even the realization of announced projects are highly uncertain; and iii)~once in operation, their power consumption exhibits distinctive temporal dynamics driven by workload characteristics. This section formalizes two core concepts used throughout the paper: \emph{uncertainty} (Section~\ref{subsec:uncertainty}) and \emph{multi-timescale variation} (Section~\ref{subsec:variation}). A spatial concentration metric is introduced to contextualize why these issues matter for grid planning. We also use cryptocurrency mining as a comparative baseline: crypto and AI share electrical infrastructure and compute hardware, but differ in economic drivers, operational flexibility, and long-horizon predictability.

\subsubsection{Spatial Concentration}

Emerging compute loads exhibit pronounced \emph{uneven spatial distribution}: large capacity expansions often concentrate at a small number of substations rather than spread across multiple feeders. To quantify this concentration, we define an area load intensity metric:
\begin{equation}
I_{\mathrm{area}}(\mathcal{A})
\;:=\;
\frac{\sum_{j \in \mathcal{D}(\mathcal{A})} P^{\mathrm{conn}}_j}{|\mathcal{A}|}
\quad
\left[\mathrm{MW/mi^2}\right],
\label{eq:areal_intensity}
\end{equation}
where $\mathcal{A}$ is a geographic region (e.g., a county, utility planning zone, or buffer around a substation), $|\mathcal{A}|$ is its land area, and $\mathcal{D}(\mathcal{A})$ is the set of large-load facilities within $\mathcal{A}$, each with interconnection capacity $P^{\mathrm{conn}}_j$.%
\footnote{In addition to AI data centers and cryptocurrency mining, several other emerging load categories exhibit similarly concentrated spatial footprints and share many of the planning challenges discussed here. These loads include Megawatt Charging Systems for freight trucks (1--3.75~MW per charger; station-level loads of 20~MW or more) \cite{nrel_electric_trucking_mcs}, industrial-scale green-hydrogen electrolyzers (100--700~MW projects operational, under construction, or at final investment decision) \cite{iea_ghr_2024}, and semiconductor fabs (roughly 200~MW for a single leading-edge fab, exceeding 1~GW for full multi-fab campus build-out) \cite{csis2024fabs}.}

\begin{table}[t]
\centering
\caption{Representative Areal Load Intensities}
\label{tab:areal_intensity}
\small
\begin{tabular}{lc}
\toprule
\textbf{Context} & \textbf{Intensity (MW/mi$^2$)} \\
\midrule
U.S.\ nationwide average & $\approx 0.125$ \\
Suburban residential & 5--20 \\
New York City (citywide avg.) & $\approx 36$ \\
Mid-sized city downtown core & 50--200 \\
Large-scale AI/DC campus & 500--1,000 \\
\;\;e.g., Stargate, TX (1.2~GW) & $\approx 700$ \\
\;\;e.g., Meta Hyperion (5~GW planned) & $\approx 1{,}000$ \\
Crypto mining campus & 1,000--2,000 \\
Manhattan (peak) & up to 2,000 \\
Ultra-dense AI data center & 2,000--3,000 \\
\bottomrule
\end{tabular}
\par\vspace{3pt}
\parbox{\linewidth}{\footnotesize Note: the U.S.\ row is an average power density; other rows are on a peak-demand or connected-capacity basis. Sources and derivations are given in the text; ranges without sources cited in the text are authors' estimates.}
\end{table}

Table~\ref{tab:areal_intensity} places compute-driven load intensities in perspective. The U.S.\ nationwide average is approximately $0.125~\mathrm{MW/mi^2}$, derived from 2023 retail sales of $3{,}861$~billion~kWh \cite{eia_electricity_sales_2023} spread over $3{,}532{,}316~\mathrm{mi}^2$ of land area \cite{census_us_land_area}. Typical suburban residential areas fall in the 5--20~$\mathrm{MW/mi^2}$ range, and New York City as a whole averages roughly 36~$\mathrm{MW/mi^2}$ (2024 forecast summer peak $\approx$11,000~MW~\cite{nyiso_gold_book_2024} across $\approx$300~$\mathrm{mi^2}$ of land area~\cite{census_nyc_land_area}).

The latest generation of AI-focused data centers exhibits substantially higher densities than traditional hyperscale facilities. Vantage~Data~Centers' Frontier campus in Texas is designed for 1.4~GW of IT load on 1,200~acres, implying a campus-average intensity near $750~\mathrm{MW/mi^2}$ \cite{vantage_frontier_2025}; smaller-footprint AI facilities with liquid-cooled, ultra-high-density racks ($\ge 250$~kW per rack~\cite{vantage_frontier_2025}) can exceed $2{,}000~\mathrm{MW/mi^2}$ (authors' extrapolation; industry surveys report rack densities above 100--200~kW becoming increasingly common~\cite{datacenterknowledge_2026_trends}). Cryptocurrency mining facilities occupy a similar range: Riot~Platforms' Corsicana campus delivers 400~MW (Phase~1) on approximately 265~acres ($\approx 1{,}000~\mathrm{MW/mi^2}$), with a planned build-out to 1~GW \cite{RiotCorsicana2024}.
Two flagship projects under construction illustrate these campus-scale intensities and the contrasting supply arrangements they induce. Meta's Hyperion campus in Richland Parish, Louisiana---announced in December 2024 at 2~GW on 2{,}250~acres and expanded in 2026 to 5~GW of compute capacity on roughly 3{,}200~acres ($\approx 1{,}000~\mathrm{MW/mi^2}$)---will be served by the vertically integrated utility Entergy Louisiana through dedicated new resources approved by the Louisiana Public Service Commission: three combined-cycle units totaling approximately 2.26~GW and up to 1.5~GW of solar, with Meta contractually covering their revenue requirement \cite{meta2026hyperion,entergy2025lpsc}. By contrast, in ERCOT's energy-only market, the OpenAI/Crusoe Stargate campus in Abilene, Texas (1.2~GW on approximately 1{,}100~acres; $\approx 700~\mathrm{MW/mi^2}$) pairs its grid interconnection with developer-procured behind-the-meter generation, including roughly 360~MW of permitted on-site gas turbines \cite{crusoe2025abilene,gevernova2025crusoe}.
For comparison, Con~Edison reports that portions of Manhattan reach load densities up to $2{,}000~\mathrm{MW/mi^2}$\footnote{See \cite{coned_2015_load_density}: EIA Energy Conference presentation (Energy Storage Activities), slide reporting Manhattan load density.}, but that urban load is distributed across many end users and feeders, whereas a data-center or mining campus of comparable intensity concentrates demand at one or a few interconnection points, amplifying both planning uncertainty and operational variability at specific grid locations. National Laboratory of the Rockies (NLR, formerly NREL)'s \emph{Speed-to-Power} data viewer provided a geospatial basis to evaluate~\eqref{eq:areal_intensity} by aggregating mapped data-center demand with infrastructure attributes \cite{nrel_speed_to_power_viewer,doe_speed_to_power}.

\subsubsection{Uncertainty and Variation}

Spatial concentration sets the stage for the two core themes developed below. Because large loads cluster geographically, forecast errors in siting and energization timing do not average out. They translate directly into localized infrastructure stress (substations, feeders, import interfaces) and, in some cases, zonal adequacy and deliverability risks. Section~\ref{subsec:uncertainty} examines the sources and implications of this \emph{uncertainty}. Once a facility is operational, its workload-driven power profile can differ markedly from traditional industrial loads; Section~\ref{subsec:variation} characterizes this inter-temporal variability across timescales from sub-seconds to years, linking workload signatures (training vs. inference) to practical metrics.

Fig.~\ref{sec2overview} summarizes how these two properties manifest across time. Uncertainty is most consequential at longer horizons (typically months to decades) because it determines whether planners procure sufficient firm capacity and whether transmission and substation reinforcements are initiated early enough to meet lead times. In this regime, the dominant risks are commitment errors (over-build versus under-build) and stranded-cost exposure when costs cannot be shifted to other customers, which can lead to project delays or downsizing. These risks motivate the planning and adequacy focus of this paper.

By contrast, variation is an operational timescale phenomenon layered on top of concentrated spatial footprints. Power-system operations and market designs already manage routine variability (load following, regulation, contingency reserves), but AI data centers can introduce unfamiliar patterns and amplitudes of change. The most acute gaps tend to arise in the \emph{sub-second to minutes} regime, where power-electronic interactions, protection/ride-through behavior, and fast ramps from coordinated workload scheduling can stress stability and operating security beyond what existing assumptions and products were designed to cover. On-site UPS and battery storage can partially mitigate ramps and provide ride-through support (Sections~\ref{sec:security_variation}--\ref{sec:affordability_variation}), but deployment is not universal, energy and duration are limited, and phenomena such as harmonic interactions and protection coordination often remain system-level concerns that behind-the-meter assets alone do not fully address. Accordingly, our variation-focused discussion emphasizes sub-second dynamics relevant to stability, and seconds-to-minutes ramps relevant to reserves, congestion management, and flexibility procurement, while also noting that longer-term variability (hours to seasons) feeds back into planning through peak coincidence and capacity accreditation.

\subsection{Uncertainty: Forecasting Under Rapid Technical and Policy Change}
\label{subsec:uncertainty}

AI data center demand is unusually difficult to forecast because it is shaped by rapidly moving and interacting factors that are largely \emph{exogenous} to traditional load forecasting models: model and hardware innovation, competitive ``AI race'' dynamics, supply-chain bottlenecks, and evolving policy and geopolitical conditions. Efficiency breakthroughs can \emph{lower} energy per unit of service (e.g., less compute per token or better utilization), while simultaneously \emph{accelerating} adoption and shifting where workloads run (cloud, colocation, on-prem, edge). The net effect is often an \emph{expanded scenario envelope}, not a narrowing one, because both demand growth and intensity can change direction quickly depending on technical and business trajectories.

This uncertainty differs from many conventional industrial loads, where demand is more tightly coupled to local economic activity, permitting timelines, and relatively stable process technologies. It also differs from cryptocurrency mining in an important long-horizon sense: while crypto mining can be highly price-responsive in the short run, protocol-level issuance rules and supply caps are set far in advance, providing a structurally different long-term context for baseline adoption trajectories. For power systems, the forecasting challenge is therefore not only the magnitude of AI load, but also its \emph{conditionality}: which projects are real, when they will energize, how quickly they will ramp, and how firm or interruptible demand will be under different grid and price conditions.

\begin{figure}
\centering 
\includegraphics[width=\columnwidth]{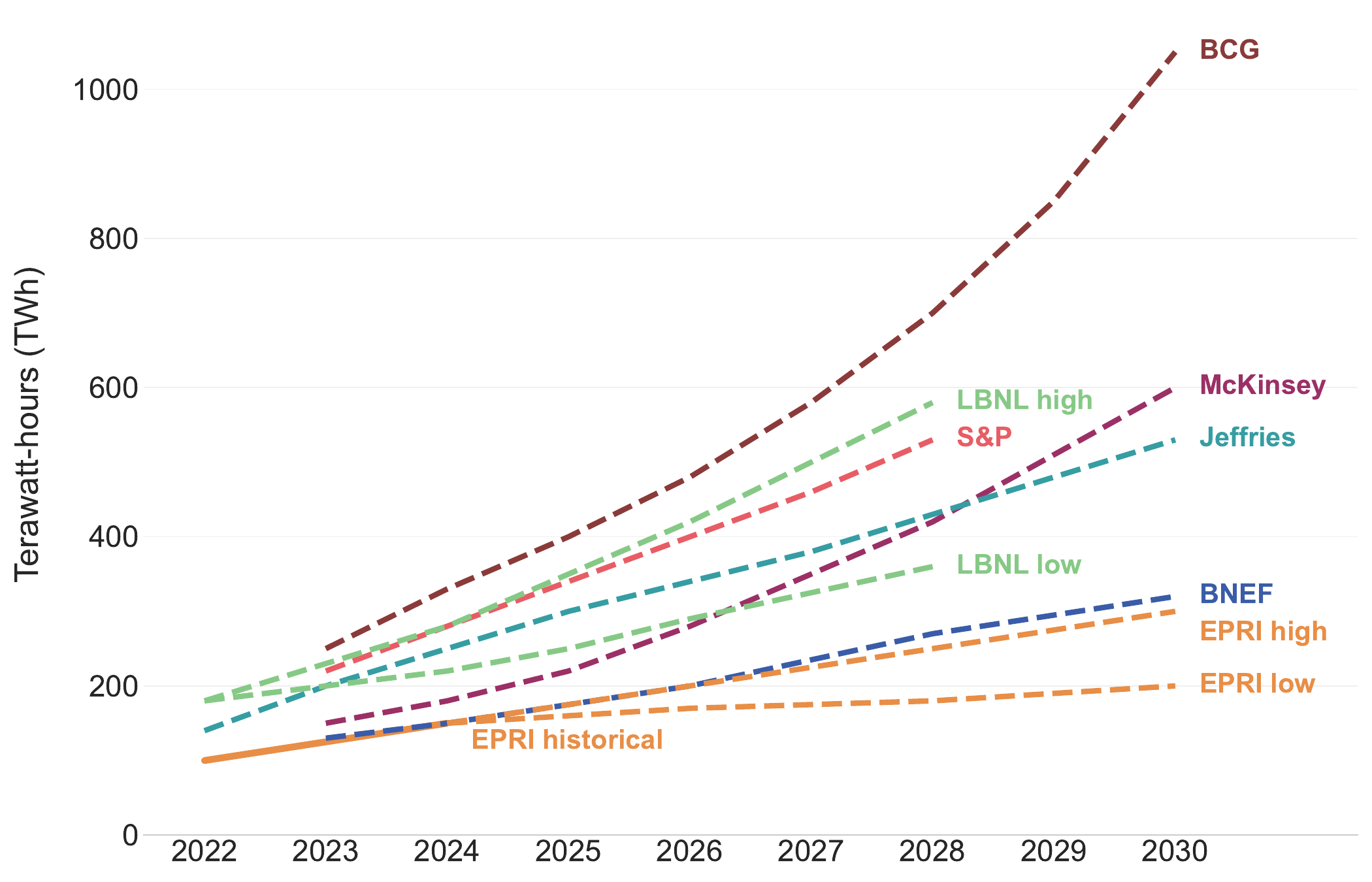} 
\caption{Published projections of U.S. data-center electricity consumption through 2030 from major forecasting organizations~\cite{WRI_2025_USDataCenterElectricityDemand}. The trajectories and reported high--low ranges reflect different assumptions about load growth and infrastructure deployment; their spread at 2030 illustrates forecast uncertainty rather than a statistical confidence interval.} 
\label{datacenterforecast1} 
\end{figure}

A practical consequence is that credible organizations can arrive at materially different views of 2030 demand even when they agree that growth will be rapid. This divergence reflects different assumptions about AI adoption and model efficiency, the share of workloads that migrate to hyperscale facilities, siting constraints and interconnection timelines, and the availability of generation and delivery infrastructure. Fig.~\ref{datacenterforecast1} illustrates this dispersion: scenario ranges published by major analysts and research organizations span multiple hundreds of TWh by 2030, implying that the standard deviation of 3+ year forecasts is substantially larger than what planners typically face for conventional industrial loads.

\begin{figure}
\centering 
\includegraphics[width=\columnwidth]{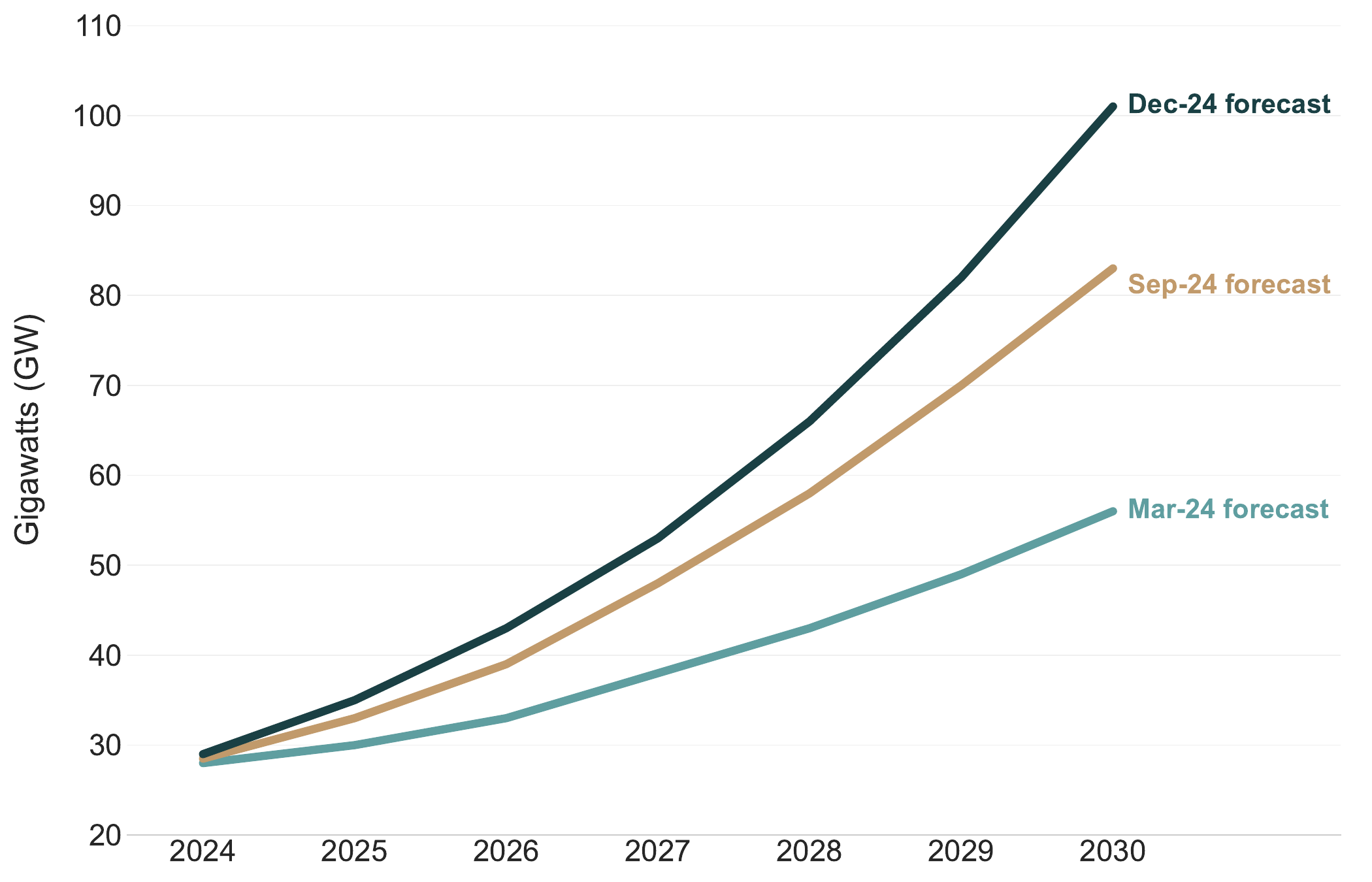} 
\caption{Successive McKinsey forecast vintages for U.S. data-center capacity through 2030~\cite{McKinsey_2024_AIPowerExpandingDataCenterCapacity}. The March, September, and December 2024 curves are alternative publication vintages over the same forecast horizon; their upward shifts show revisions as new information became available, not realized capacity growth between the publication dates.} 
\label{datacenterforecast2} 
\end{figure}

Moreover, forecasting is itself a moving target: estimates are repeatedly revised as new information arrives about interconnection queues, procurement strategies, hardware supply, and policy constraints. In some cases, published outlooks have shifted significantly within months as assumptions about build rates, load realization, and infrastructure bottlenecks are updated. Fig.~\ref{datacenterforecast2} provides an example of this ``forecast drift'' over time. For system planners and regulators, the implication is methodological: forecasts of data center demand should be updated iteratively as new information becomes available, with explicit milestones used to distinguish exploratory inquiries from committed projects and to revise scenarios as projects advance toward construction and energization.

Another related and often underappreciated source of forecast instability is the risk of \emph{double counting} large-load projects across jurisdictions and interconnection queues. Data center developers have strong incentives to explore multiple siting and procurement options in parallel: initiating service inquiries, interconnection requests, or preliminary regulatory reviews with more than one utility or transmission provider before committing to a single location. From the developer's perspective, such redundancy is rational risk management in the face of uncertain power availability, pricing, and permitting timelines. From the system planner's perspective, however, these parallel inquiries can inflate aggregate load forecasts if each request is treated as an independent, additive project. This phenomenon closely mirrors the well-documented experience on the supply side, where wind, solar, and storage developers routinely enter multiple interconnection queues, and historically less than a quarter of queued projects ultimately reach commercial operation. Without explicit mechanisms to identify mutually exclusive alternatives, load forecasts can systematically overstate realized demand and bias resource adequacy assessments toward over-procurement.

\subsection{Inter-temporal Variability and Multi-Scale Load Dynamics}\label{subsec:variation}

\subsubsection{Workload-driven variability: training vs. inference.}\label{subsec:train_vs_infer}

Even after a data center's nameplate capacity is known, its inter-temporal variation can be atypical for large loads. \emph{Training} workloads synchronize iterative computation across many accelerators; that synchronization produces coherent, step-like aggregate swings (cause) that propagate through the facility power-delivery chain (mechanism), and appear as fast ramps at the grid interface (impact). \emph{Inference} workloads are constrained by latency and driven by bursty user demand; those demand bursts (cause) concentrate power changes into diurnal and event-driven patterns (mechanism), elevating congestion exposure and coincident-peak contribution even when annual energy is unchanged (impact). These behaviors span milliseconds to years: sub-second excursions interact with voltage control, protection, and power-quality compliance, while minute-scale ramps stress reserves, congestion management, and dispatch feasibility. ``Average MW'' therefore fails to describe the operational and planning risk profile in the AI era.

\subsubsection{A timescale crosswalk from workload signatures to grid observables.}\label{subsec:timescale_crosswalk}

We adopt a two-level timescale hierarchy that links i)~representative phenomena, ii)~primary facility- and grid-side components (``actors''), and iii)~quantitative metrics used later in Sections~\ref{sec:adequacy_uncertainty}--\ref{sec:affordability_variation}.%
\footnote{This hierarchy is a \emph{modeling crosswalk}, not a claim that any single timescale alone determines system risk. The intent is to clarify which physics and controls are dominant, what the observables are, and which tools (electromagnetic transient (EMT), phasor-domain dynamics, production-cost models, or planning models) are appropriate.}
The fast-dynamics layer ($\mu$s--60\,s) captures converter-rich behavior and control interactions that determine \emph{survivability} and compliance (power quality, ride-through, and sub-minute regulation behavior)~\cite{NERC2025WhitePaper,WECC2025LargeLoads}. The operational layer (1\,min--24\,h) captures dispatch- and market-relevant variability (net-load ramps, reserve needs, congestion exposure, and diurnal/event-driven structure)~\cite{Wang2019,DOE2025DCFlex}.

Fast-timescale phenomena impose \emph{hard} feasibility constraints that define an operating envelope; within that envelope, operational coordination allocates scarce flexibility (or reveals its absence), shaping near-term congestion and curtailment risk~\cite{NERC2025Alert}.

\paragraph{Fast dynamics ($\mu$s--60\,s): feasibility and compliance (overview)}\label{subsec:fast_dynamics_overview}
At this layer, the central questions are ``\emph{can the system ride through and remain compliant?}'' rather than ``\emph{what is the marginal price?}''~\cite{ERCOT2025Stability}. Converter-dominated power delivery (server power supply units (PSUs) and voltage regulator modules (VRMs), rectifier front ends, harmonic filtering, and UPS and battery energy storage system (BESS) interfaces) makes data centers salient, and synchronized \emph{training} can compress many server-level transitions into coherent steps at the point of interconnection~\cite{Chen2025}. Representative observables include harmonic/interharmonic distortion and flicker (e.g., IEC~61000-4-7/4-15), voltage unbalance, sag/swell tolerance (e.g., ITI/CBEMA), RoCoF and frequency nadir, and sub-minute ramp rates relevant to regulation behavior~\cite{IEC61000-4-7,IEC61000-4-15,ITI-CBEMA-2000,ieee_519_2022,NERC2025Incident}. Appropriate tools include EMT simulation or impedance-based screening for interaction risks, and validated dynamic models of UPS and IT power controls~\cite{Samanta2024}.

\paragraph{Operations (1\,min--24\,h): dispatch and market-relevant variability (overview)}\label{subsec:operations_overview}
Minute-to-day variability is where feasibility constraints translate into economic outcomes: commitment feasibility, reserve sufficiency, congestion, scarcity pricing, and curtailment~\cite{Ghatikar2010}. Representative metrics include MW/min ramp rates (and rebound), reserve sufficiency margins, binding-constraint indicators, scarcity adders, and locational marginal price (LMP) statistics (mean/variance/spikes), along with response-time and performance measures for demand response participation~\cite{ERCOTDR2022}. Mechanisms that matter include workload orchestration (shifting/deferral), use of UPS/BESS as short-duration buffers, and thermal/cooling coupling that shapes sustained loading during hot conditions~\cite{Paananen2023Electrification,Liu2020TES,Winick2025UTES}.

\subsubsection{Spectral characteristics and multi-scale validation.}\label{subsec:spectral_validation}
\begin{figure*}[!t]
    \centering
    \includegraphics[width=\textwidth]{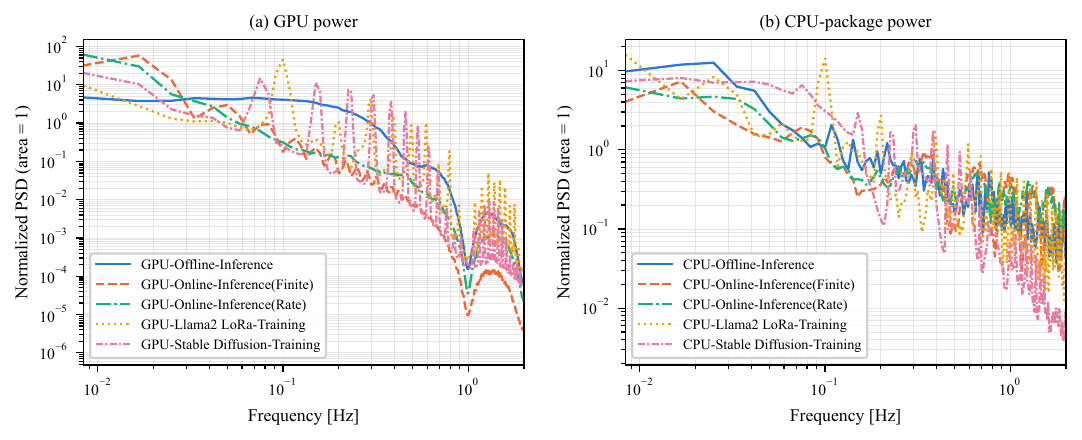}
    \caption{Area-normalized Welch power spectral densities (120\,s Hann window, 50\% overlap) for five representative single-node AI workload traces from the NLR Dataset of Generative AI Workload Power Profiles~\cite{NLR2026Dataset,vercellino2026measurement}, shown for (a) GPU power and (b) CPU-package power.}
\label{fig:nlr_psd_spectrum}
\end{figure*}

We validate this multi-scale structure directly using the NLR Dataset of Generative AI Workload Power Profiles~\cite{NLR2026Dataset,vercellino2026measurement}: area-normalized Welch power spectral densities (PSDs) for the GPU and CPU-package channels for five representative training and inference traces (Fig.~\ref{fig:nlr_psd_spectrum}). Training workloads show clean harmonic combs at a $\sim$10--13\,s step cadence, recovering the training-step period directly from the power spectrum; offline and online-rate inference share a distinct $\sim$0.75\,s peak suggestive of inference-engine-level periodicity, and GPU spectra are markedly more tonal than the broadband CPU-package spectra throughout.

\begin{figure*}[!t]
    \centering
    \includegraphics[width=\textwidth]{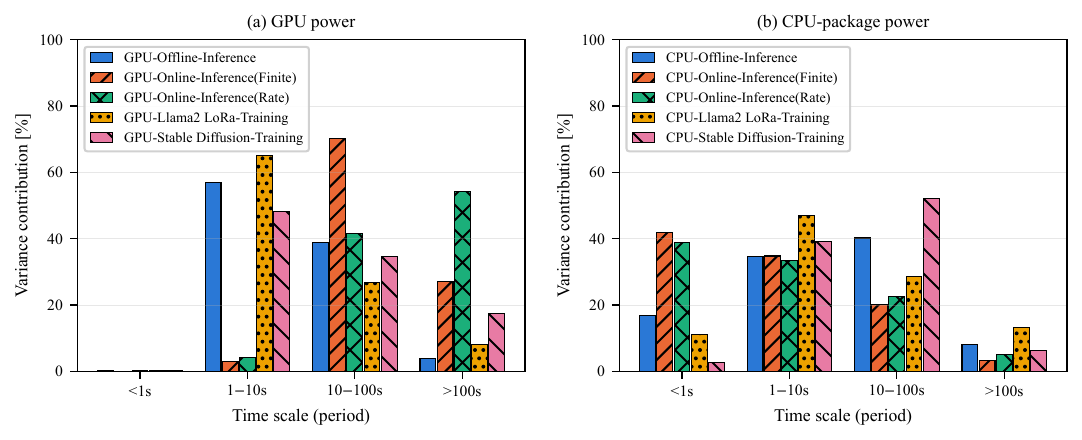}
    \caption{Variance decomposition by time scale ($<$1\,s, 1--10\,s, 10--100\,s, $>$100\,s, the last bounded by the 120\,s Welch-window resolution) for the same five traces, shown for (a) GPU power and (b) CPU-package power.}
\label{fig:nlr_timescale_variance}
\end{figure*}

Decomposing variance by time scale (Fig.~\ref{fig:nlr_timescale_variance}) shows that the dominant band depends on the workload: training and offline inference concentrate variance in the 1--10\,s band, whereas the online inference variants shift toward 10--100\,s and above. Sub-second content accounts for under 1\% of GPU variance in every trace, although the CPU-package channel carries substantially more sub-second content for the online variants. These are properties of the five measured traces rather than universal characteristics of AI workloads, and the spectral structure should be expected to differ with model, serving configuration, and hardware. The point we draw from them is methodological: the relevant structure of these loads is workload-dependent and lives at time scales that peak-demand descriptors do not express, which is also what identifies where demand-flexibility and dynamic voltage and frequency scaling (DVFS) interventions are best matched~\cite{AlKez2021}.

\section{Reliability Under Uncertainty: System-Level Resource Adequacy}\label{sec:adequacy_uncertainty}

\begin{figure}[H]
    \centering
    \includegraphics[width=1\columnwidth]{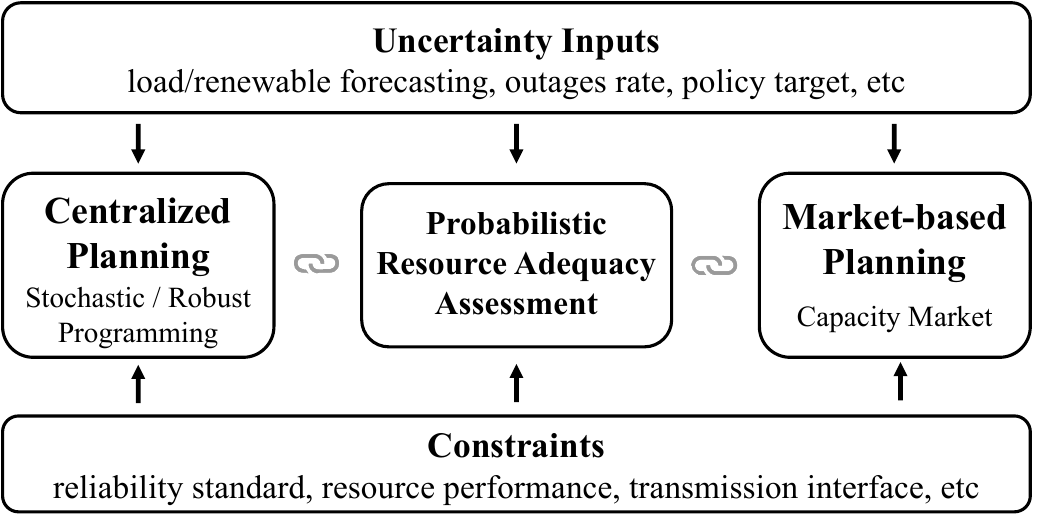}
\caption{Two institutional pathways to resource adequacy and their common analytical core. Uncertainty inputs such as load and renewable forecasts, outage rates, and policy targets enter probabilistic resource adequacy assessment under reliability, resource-performance, and transmission constraints. The resulting adequacy information supports either centralized stochastic/robust planning or market-based procurement through capacity markets.}
\label{fig:uncertainty_planning_framework}
\end{figure}

In this section, we use \emph{reliability} in the planning sense, focusing primarily on system-level \emph{resource adequacy}: whether the system has sufficient deliverable resources to serve demand across credible realizations of uncertainty. As illustrated in Fig.~\ref{fig:uncertainty_planning_framework}, there are two broad institutional pathways by which jurisdictions seek to ensure resource adequacy. The first is centralized planning, where a utility or system planner selects cost-effective investments and adequacy is enforced through planning criteria and regulatory processes \cite{koltsaklis2018state}. The second is market-based planning, where investment incentives are mediated through capacity markets with reliability requirements translated into market demand curves, performance rules, and deliverability requirements \cite{joskow2006competitive}. Importantly, these two pathways both rely on probabilistic resource adequacy assessment (RAA) to map uncertainties and constraints into quantitative risk metrics and compliance determinations \cite{tillmanns2026review}. 

An important resource option in this context is behind-the-meter (BTM) generation at the data-center site. It is useful to distinguish BTM generation that is intended to serve the data-center load on a routine basis from on-site backup generation. BTM generation may supply all or part of the facility's electricity demand as a substitute for grid-supplied generation, thereby reducing the net load for which the bulk power system must plan and maintain resource adequacy. By contrast, backup generation is primarily intended to maintain facility operation during grid outages or other emergency conditions and should not automatically be treated as a resource that offsets system-supplied demand for resource-adequacy purposes. This distinction is important because BTM supply and backup generation are sometimes conflated, even though their operating roles and implications for system planning can be substantially different.

In the following, we first briefly introduce the probabilistic RAA model and show how AI data-center growth expands the uncertainty affecting load modeling, complicating the interpretation of standard adequacy metrics. We then discuss how these uncertainties challenge centralized planning by increasing the value of staged commitments and tightening the coupling between generation expansion and transmission deliverability. Finally, we turn to market-based planning and explain how data-center uncertainty stresses capacity market design, including demand-curve calibration, deliverability requirements, and performance incentives. 

\subsection{Probabilistic Resource Adequacy Assessment}

\subsubsection{System Model in RAA} 

Regardless of whether adequacy is pursued through centralized planning or market-based mechanisms, compliance is often evaluated by Monte Carlo simulation tools rather than being explicitly embedded in the optimization program or market-clearing process. Conventional RAA models were designed around the forced-outage behavior of thermal generators, simulating unit-level outages using technology-specific outage rates and repair-time distributions~\cite{allan2013reliability}. Modern RAA simulators, such as GE-MARS~\cite{GEMARS} and PRAS \cite{stephen2024probabilistic},  extend this framework to include variable renewable resources and storage: renewable availability is typically represented using historical (or weather-simulated) production profiles, while storage is modeled through an assumed dispatch policy that maps net-load conditions into charging and discharging decisions.

Let $\mathcal{T}$ denote the set of all hours in a year (or the study periods of interest), $\mathcal{D}$ the set of all days, and $\Omega$ the set of all possible year-long scenarios for the system under study, assumed to be finite. Following the notation in \cite{qiancredit}, we construct a probability space $(\Omega, \mathcal{F}, \mathbb{P})$, where $\mathcal{F}$ represents the $\sigma$-algebra of events on $\Omega$ and $\mathbb{P}$ is a probability measure defined on $\mathcal{F}$ satisfying standard measure-theoretic properties. Let $X: \Omega \rightarrow \mathbb{R}$ be a random variable representing our resource adequacy (RA) metric for any given scenario. For each generator $g \in \mathcal{G}$, let $x_g \in \mathbb{R}_{+}$ denote its installed capacity. In the absence of storage, and assuming that generation resources have no intertemporal operating constraints, we define $
A_{g\tau\omega} : \mathcal{T} \times \Omega \rightarrow [0,1]$ as a stochastic process representing generator $g$'s availability, where $A_{g\tau\omega}$ denotes its realization in hour $\tau$ of scenario $\omega$. Then, the total available capacity is the stochastic process on $\mathcal{T}\times\Omega\to\mathbb{R}_{+}$ defined as
\begin{equation}
\hat{x}_{\tau\omega} = \sum_{g \in \mathcal{G}} x_g A_{g\tau\omega}.
\end{equation}
The system load profile is typically sampled from a set of representative scenarios, with $L_{\tau\omega}$ denoting load in hour $\tau$ of scenario $\omega$. For notational compactness, we define the vectors of hourly values within scenario $\omega$ as
\begin{equation}
\boldsymbol{\hat{x}}_\omega := (\hat{x}_{\tau\omega})_{\tau \in \mathcal{T}},
\quad
\boldsymbol{L}_\omega := (L_{\tau\omega})_{\tau \in \mathcal{T}},
\end{equation}
representing total available capacity and system load, respectively.

\subsubsection{Resource Adequacy Metrics}
For semantic clarity, we distinguish three related but distinct concepts: the \emph{RA metric}, the \emph{risk measure}, and the \emph{reliability standard}. The RA metric specifies how shortage events are quantified and aggregated within a single assessment period, such as Unserved Energy (UE), Loss of Load Hours (LOLH), and Loss of Load Days (LOLD). The risk measure defines how these per-scenario outcomes are aggregated across multiple stochastic realizations (e.g., expectation or percentile). Finally, the reliability standard specifies the target or benchmark value of the risk measure that the system is expected to satisfy, such as the `\emph{one day in ten years}' LOLD criterion commonly used in North America since the 1960s \cite{stephen2022clarifying}. We define the RA metric for each scenario $\omega$ as a functional
\begin{equation}
M: \mathbb{R}_+^{|\mathcal{T}|} \times \mathbb{R}_+^{|\mathcal{T}|} \to \mathbb{R}_+,
\quad
M(\boldsymbol{\hat{x}}_\omega, \boldsymbol{L}_\omega),
\end{equation}
which aggregates unserved load over time according to the chosen reliability measure. For example, LOLH can be represented as: 
\begin{equation} \label{lole}
M_{\text{LOLH}}(\boldsymbol{\hat{x}}_\omega, \boldsymbol{L}_\omega)
= \sum_{\tau \in \mathcal{T}} \mathbbm{1}\Big\{ (L_{\tau\omega} - \hat{x}_{\tau\omega}) > 0 \Big\}.
\end{equation}
The system-level RA measure is defined by aggregating the per-scenario metric across the probability space:
\begin{equation} \label{metric}
\mathcal{M}(\boldsymbol{\hat{x}}, \boldsymbol{L})
:= \Phi_\omega \big[\, M(\boldsymbol{\hat{x}}_\omega, \boldsymbol{L}_\omega) \,\big],
\end{equation}
where $\Phi_\omega[\cdot]$ is a risk operator, typically expectation.

{
\subsubsection{Effective Load Carrying Capability}

Effective Load Carrying Capability (ELCC) provides a reliability-based measure of the capacity contribution of a resource. In words, the (marginal) ELCC of a candidate unit is the largest constant load increase that can be served after adding the unit while maintaining the same system-level resource-adequacy risk.

Consider a candidate resource $g$ with nameplate capacity $\Delta x>0$ and scenario-dependent availability
$\boldsymbol{A}_{g}:=(A_{g\tau\omega})_{\tau\in\mathcal{T},\,\omega\in\Omega}$.
Using the system-level RA measure $\mathcal{M}$ defined above, its marginal ELCC is defined as
\begin{equation}
\begin{split}
L_c^{\mathrm{ELCC}}:=\max_{L_c\geq 0}\Big\{ L_c:\;
&\mathcal{M}\left(\boldsymbol{\hat{x}}+\Delta x\,\boldsymbol{A}_{g},\,\boldsymbol{L}+L_c\mathbf{1}\right)\\
&\leq\mathcal{M}\left(\boldsymbol{\hat{x}},\boldsymbol{L}\right)\Big\},
\end{split}
\label{eq:elcc-implicit}
\end{equation}
where $\mathbf{1}$ is the all-ones vector, so that $L_c\mathbf{1}$ is a constant load increase across all study periods and scenarios. Equivalently, at the limiting value, the additional resource and additional load leave system reliability unchanged. The corresponding ELCC capacity accreditation is
\begin{equation}
\alpha_g^{\mathrm{ELCC}}
:=
\frac{L_c^{\mathrm{ELCC}}}{\Delta x},
\qquad
\alpha_g^{\mathrm{ELCC}}\in[0,1].
\label{eq:elcc-credit}
\end{equation}

ELCC therefore provides an important link between probabilistic RAA and planning or market decisions. In centralized planning, ELCC-based capacity accreditation converts heterogeneous resources into comparable firm-capacity contributions and thereby informs whether a portfolio satisfies a planning reserve-margin (PRM) requirement. In capacity markets, the same principle determines the quantity of accredited capacity that a resource may offer or receive compensation for, effectively converting resources with different availability characteristics into a common reliability product. Because ELCC depends on the existing resource portfolio, load profile, and joint availability conditions, it is inherently system- and portfolio-dependent rather than a fixed rating.
}

\subsubsection{Impact of Data Center and Increasing Uncertainty in Resource Adequacy Study}

Rapid expansion of data centers affects probabilistic RAA \emph{primarily} by introducing more uncertainty in the load profile $\boldsymbol{L}_\omega$ and by strengthening common-mode stress conditions under which high demand coincides with degraded resource availability \cite{mantegna2025electric}. In the notation above, the key modeling implications are: i)~enlarging the scenario set $\Omega$ to represent uncertainty in data-center timing and scale, and ii)~representing an uncertain operational characteristic that reduces \emph{effective} load during scarcity, which decomposes $L_{\tau\omega}$ into baseline load plus an incremental data-center component \cite{colangelo2025ai}. Taken together, these changes tend to increase the variance and tail risk of shortage outcomes, and also impact the interpretation of standard adequacy metrics.

\paragraph{Data Center Demand and Flexibility}

Recent work emphasizes that modern resource adequacy is no longer a purely supply-side problem with fixed demand; flexible demand can materially affect both the frequency and severity of load shedding, and large loads such as data centers introduce additional uncertainty because their flexibility is often imperfectly known ex ante \cite{mantegna2025electric,sun2022insights}. In our notation, a compact way to represent data-center-driven uncertainty is to decompose hourly load into a baseline component and a data-center component:
\begin{equation}
L_{\tau\omega} \;=\; L^{\mathrm{base}}_{\tau\omega} \;+\; L^{\mathrm{DC}}_{\tau\omega},
\qquad
L^{\mathrm{DC}}_{\tau\omega} \;=\; K_\omega\, s_{\tau\omega},
\end{equation}
where $K_\omega$ captures the realized scale (MW) of data-center additions in scenario $\omega$, and $s_{\tau\omega}\in[0,1]$ summarizes the energization and utilization ramp. This representation makes explicit why expanding the scenario set $\Omega$ is often necessary: uncertainty in $(K_\omega,s_{\tau\omega})$ can dominate the upper tail of $L_{\tau\omega}$ even when $L^{\mathrm{base}}_{\tau\omega}$ is sampled from representative weather years.

To reflect the modern distinction between \emph{static} demand and \emph{flexible} demand in adequacy studies, one model is to treat part of the data center load as potentially controllable during system stress \cite{kuna2024beyond}. Let $r_{\tau\omega}\ge 0$ denote the realized voluntary reduction (e.g., contracted demand response, managed curtailment, or behind-the-meter supply netting) available in hour $\tau$ and scenario $\omega$. Then the system-served (effective) load becomes
\begin{equation}
L^{\mathrm{eff}}_{\tau\omega} \;=\; L_{\tau\omega} - r_{\tau\omega},
\qquad
0 \le r_{\tau\omega} \le \bar r_{\tau\omega},
\end{equation}
where $\bar r_{\tau\omega}$ is the flexibility envelope. In probabilistic RAA, the value of flexibility is driven mainly by its \emph{coincidence} with scarcity hours rather than total curtailed energy. Consequently, uncertainty in the availability of $r_{\tau\omega}$ should be reflected in the scenario set $\Omega$ whenever data center flexibility is material.

\paragraph{Metric Sensitivity} 

Because data-center growth often introduces ``lumpy'' additions and steeper peak conditions, it can shift the shortage profile from many small deficits to fewer but larger events (or vice versa), which affects the relative informativeness of common per-scenario metrics \cite{stephen2022clarifying}. For instance, severity-sensitive metrics such as unserved energy,
\begin{equation}
M_{\mathrm{UE}}(\omega)=\sum_{\tau\in\mathcal{T}} \big(L^{\mathrm{eff}}_{\tau\omega}-\hat{x}_{\tau\omega}\big)_+,
\end{equation}
can distinguish plans that have similar event counts but very different deficit magnitudes, whereas frequency-based metrics such as LOLH \eqref{lole}
primarily reflect how often scarcity occurs. Under increased uncertainty, tail-aware risk aggregation $\Phi_\omega[\cdot]$ (e.g., conditional value at risk, CVaR) can become more relevant when adequacy concerns are driven by adverse combinations of high load growth and stressed system conditions.

\paragraph{Co-Location}

The term \emph{co-location} admits multiple interpretations across jurisdictions and regulatory contexts. For example, under PJM's proposed definition, a co-located load refers to end-use customer load that is physically connected to the facilities of an existing or planned generator on the generator's side of the point of interconnection (POI) to the transmission system \cite{FacontiFrankHoweRobbinsThompson2026}. Data Center co-location with on-site generation, storage, or firm imports can reduce the contribution of data centers to system risk by decreasing the load seen by the bulk system during stressed hours \cite{li2025risk}. In the RAA abstraction, this effect is naturally captured through the same $r_{\tau\omega}$ term (netting behind-the-meter supply from load) and, crucially, by allowing $(r_{\tau\omega},L_{\tau\omega})$ and the system availability process $(A_{g\tau\omega})$ to be correlated under common drivers (e.g., weather and fuel constraints). This correlation structure poses a new challenge when data-center growth is rapid and geographically concentrated.

\subsection{Centralized Planning}
\label{subsec:no_capacity_market}

In vertically integrated systems, resource adequacy is primarily ensured through integrated resource planning, utility procurement, and regulatory approval of long-lived investments. Under uncertain data center growth, the core challenge is not simply selecting an optimal portfolio for a given forecast, but sequencing irreversible investment decisions in the presence of evolving information. Over-commitment risks stranded assets and rate impacts, while under-commitment increases loss-of-load risk and reliance on emergency measures.

\subsubsection{Formulation of Generation Expansion Planning Under Uncertainty}
\label{subsubsec:gep_formulation} 

Generation expansion planning (GEP) determines long-lived capacity investments while ensuring that the resulting system can be operated feasibly and economically. A standard modeling backbone is a two-level (investment--operation) structure: a \emph{first stage} commits capacity expansion, and a \emph{second stage} performs system operation (e.g., economic dispatch) subject to operational and network constraints. Deterministic and stochastic variants have a long history in the GEP literature \cite{Manne1961,freidenfelds1980capacity,BeanHigleSmith1992,BienstockShapiro1988,JinEtAl2011,sun2021robust}. This framework also provides a natural entry point for discussing how AI data-center uncertainty shifts attention from static planning toward more adaptive (but computationally harder) multi-stage formulations \cite{BermanGanzWagner1994,ahmed2003multi,ahmed2003approximation,zou2018partially}.

Let $x_c$ denote the installed (or incremental) capacity decision for technology $c$ (MW), with unit investment cost $\mathrm{Inv}_c$ and upper bound $\bar{x}_c$.
Let $\omega\in\Omega$ index uncertainty scenarios with probability $p^\omega$.
Let $y^\omega$ denote operating decisions in scenario $\omega$ (dispatch, flows, reserves, etc.).
Let $d^\omega$ denote scenario-dependent net load (demand net of variable supply if modeled that way).
Let $X(d^\omega)$ denote the operational feasibility set, such as power balance, generator limits, ramping, transmission constraints, etc.

\paragraph{Deterministic Two-stage Formulation}
The deterministic formulation replaces uncertainty with a single load forecast $d=(d_t)_{t\in\mathcal{T}}$ and chooses capacity investments $x$ to minimize investment plus operating cost:
\begin{equation}
\label{eq:det_stage1}
\begin{aligned}
\min_{x}\quad & \sum_{c} \mathrm{Inv}_c\, x_c + Q(x)\\
\text{s.t.}\quad & 0 \le x_c \le \bar{x}_c,\quad \forall c,
\end{aligned}
\end{equation}
where the second-stage operating cost is given by
\begin{equation}
\label{eq:det_stage2}
\begin{aligned}
Q(x)=\min_{y}\quad & \sum_{t\in\mathcal{T}}\sum_{g} C^E_{gt} P^E_{gt}\\
\text{s.t.}\quad & (y,x)\in X(d).
\end{aligned}
\end{equation}
Here $y$ collects operational decisions, $P^E_{gt}$ denotes generation output, and $X(d)$ represents standard operational and network constraints under the deterministic load forecast $d$.

\paragraph{Stochastic Two-stage Formulation}
To represent uncertainty explicitly, the deterministic operating term is replaced by an expectation over scenarios:
\begin{equation}
\label{eq:stoch_stage1}
\begin{aligned}
\min_{x}\quad & \sum_{c} \mathrm{Inv}_c\, x_c
+ \sum_{\omega\in\Omega} p^\omega\, Q(x,\omega)\\
\text{s.t.}\quad & 0 \le x_c \le \bar{x}_c,\quad \forall c,
\end{aligned}
\end{equation}
where the second-stage operating cost in each scenario $\omega$ is given by
\begin{equation}
\label{eq:stoch_stage2}
\begin{aligned}
Q(x,\omega)=\min_{y^\omega}\quad
& \sum_{t\in\mathcal{T}}\sum_{g} C^E_{gt}\, P^{E\omega}_{gt}\\
\text{s.t.}\quad
& (y^\omega,x)\in X(d^\omega).
\end{aligned}
\end{equation}
In this ``here-and-now'' structure, capacity decisions $x$ are committed before uncertainty is realized, while operational decisions $y^\omega$ adapt to each scenario. Depending on the planner's risk preferences, the expected-cost objective can be replaced by alternative risk-aware formulations, such as robust optimization or chance-constrained planning, to hedge against adverse load realizations \cite{pourahmadi2021distributionally,yan2019scenario}.

\paragraph{Multi-stage Stochastic Optimization}

Two-stage models can under-value flexibility when uncertainty resolves gradually over multiple planning cycles, as is common with large load additions. Multi-stage stochastic GEP generalizes \eqref{eq:stoch_stage1}--\eqref{eq:stoch_stage2} by allowing capacity decisions to adapt over multiple periods as uncertainty unfolds on a scenario tree, subject to non-anticipativity \cite{BermanGanzWagner1994,ahmed2003multi}. The corresponding challenge is computational: scenario trees grow rapidly with horizon length and branching, and realistic multi-stage models (often with discrete investment decisions) can become large-scale stochastic programs that are difficult to solve and update frequently \cite{ahmed2003approximation}.

\subsubsection{Impact of Data Center Uncertainty in Centralized Planning}

Rapid growth in data center demand changes centralized planning primarily by increasing i)~the value of \emph{adaptivity} in investment decisions, ii)~the coupling between generation and transmission expansion, and iii)~the difficulty of enforcing reliability requirements \emph{endogenously} within an optimization model. Unlike ex post Monte Carlo RAA, a planning model must commit long-lived decisions before key uncertainty drivers are revealed, so the formulation must trade computational tractability against fidelity to adequacy risk.

\paragraph{Increasing Need for Multi-stage and Adaptive Planning} Large data center additions are often uneven and only partially revealed over time (e.g., as interconnection milestones are met, construction progresses, and tenant/utilization becomes firm). This information structure is not well captured by a ``here-and-now'' two-stage model, which commits all investments before uncertainty is realized. Multi-stage formulations instead allow investment decisions to adapt across planning stages as new information arrives.

The computational cost, however, rises quickly with the number of stages. Each additional stage typically requires a richer scenario representation of how uncertainty unfolds (i.e., more branching in the scenario tree) and introduces a new set of stage-dependent decision variables replicated across scenario nodes. As a result, model size can grow roughly in proportion to the number of nodes in the scenario tree, making realistic multi-stage stochastic GEP substantially harder to solve.

\paragraph{Increasing Coupling with Transmission Planning} Geographically concentrated data center growth makes deliverability constraints bind more frequently, so planning must increasingly co-optimize generation and transmission. In compact form, this expands the feasibility set from $X(d^\omega)$ to a network-aware set that includes transmission build decisions $(y^\omega,x)\in X(d^\omega)$ to $(y^\omega,x,z)\in X(d^\omega;z)$. A planner-initiated siting framework that pre-screens interconnection-ready locations under standardized operational-flexibility envelopes (firm, pause, shift) operationalizes this coupling directly: on a synthetic 2000-bus Texas system it expands the siting frontier by 9--21\% and cuts first-energization timelines from 5--8 years to 12--18 months relative to firm operation~\cite{kim2026siting}.

Transmission expansion often introduces binary investment variables and lumpy project choices, so even if the generation expansion model is linear/convex, joint planning typically becomes a mixed-integer program  \cite{gonzalez2020review}. In a multi-stage setting, integrality interacts with non-anticipativity and scenario growth, substantially increasing solve time and limiting how frequently the plan can be re-optimized as new data center information arrives.

\paragraph{Embedding Reliability Requirements into GEP} Centralized planning models often enforce adequacy through simplified proxies (reserve margins, capacity constraints, or deterministic peak requirements), while probabilistic adequacy is evaluated ex post via Monte Carlo simulation and metrics such as loss of load expectation (LOLE) and expected unserved energy (EUE). Embedding the same reliability requirement directly into the investment optimization is difficult because reliability metrics are defined over a large scenario space and depend on detailed scarcity dynamics. A reduced-form approach is to introduce an explicit shortage variable in the dispatch layer and penalize it with the value of lost load (VoLL)
or to impose an explicit risk constraint on unserved energy, consistent with the RAA notation \eqref{metric}. These embeddings make reliability endogenous, but they also highlight the core tension: higher-fidelity reliability constraints push the model toward large-scale stochastic optimization, motivating approximations, decomposition, or hybrid simulation-optimization frameworks \cite{fu2002optimization}.

\subsection{Systems With a Capacity Market}
\label{subsec:capacity_market}

Capacity markets ensure long-term resource adequacy in deregulated electricity markets by providing financial incentives for resources to commit their capacity in advance \cite{joskow2006competitive}. In these markets, the system operator procures \emph{accredited} capacity in advance of the delivery year to satisfy an administratively chosen reliability criterion, most commonly expressed as a target LOLE. Procurement is implemented through a forward auction that clears against a demand curve, which translates the reliability target and the underlying probabilistic RAA into a willingness-to-pay for capacity. 

Capacity markets act as a \emph{risk-management layer} between probabilistic adequacy modeling and decentralized investment, using forward price signals to encourage entry when scarcity risk rises and to deter overinvestment when the system is long. In practice, they coexist with heterogeneous state policies, bilateral contracting, and complementary reliability tools such as local deliverability requirements and out-of-market procurement. Table~\ref{tab:ra_by_rto_short} summarizes key differences in resource adequacy institutions across U.S.\ ISO/RTOs: some regions rely primarily on load-serving-entity (LSE) resource adequacy obligations and bilateral/self-supply arrangements, while others rely more heavily on ISO-run centralized capacity auctions. NYISO represents a hybrid case, where LSEs retain RA obligations but may satisfy them through bilateral contracts, self-supply, and NYISO-administered capacity auctions. The remainder of this subsection focuses on how data-center-driven load uncertainty challenges the design of centralized capacity markets, particularly through demand-curve calibration, deliverability constraints, and performance incentives.

\begin{table}
\centering
\small
\caption{Simplified U.S. RA constructs by ISO/RTOs}
\label{tab:ra_by_rto_short}
\begin{tabular}{lcc}
\hline
\textbf{ISO/RTO} & \textbf{LSE RA Obligation} & \textbf{Capacity Market} \\
\hline
CAISO  & Yes & No \\
ERCOT  & No  & No \\
SPP    & Yes & No \\
MISO   & Yes & Yes \\
PJM    & No  & Yes \\
NYISO  & Yes & Yes \\
ISO-NE & No  & Yes \\
\hline
\end{tabular}
\end{table}
\emph{Remark}: ERCOT relies on its operating reserve demand curve (ORDC) and contracted demand response. PJM has an ISO-level centralized capacity market and also offers the Fixed Resource Requirement (FRR) alternative, under which LSEs may self-supply or contract for capacity. NYISO is classified as having both an LSE RA obligation and a capacity market: LSEs certify bilaterally contracted or self-supplied capacity before the NYISO-administered auction process, and any residual shortfall can be met through NYISO's monthly/spot capacity market.

\subsubsection{Demand Curve Under Load Uncertainty}
\label{subsubsec:capacity_demand_curve}

Modern capacity markets have largely moved away from clearing a fixed (vertical) capacity target and toward \emph{downward-sloping} capacity demand curves \cite{byers2018capacity}. The key motivation is to explicitly represent the economic trade-off between higher procurement cost and higher reliability, and to reduce extreme price volatility that can occur when a vertical demand curve intersects a relatively inelastic supply stack near the target quantity. In practice, these curves are implemented as administratively defined demand schedules in forward auctions, anchored to a system reliability requirement and a benchmark for investment economics such as Cost of New Entry (CONE), which is used to indicate the amortized capital cost of constructing a power
plant.

While the high-level principle is shared, implementations differ materially across ISO/RTOs. For example, PJM's curve is parameterized around a region-wide reliability requirement in UCAP terms and Net CONE, and includes an explicit price-cap segment before transitioning to downward-sloping segments \cite{SpeesNewellThompson2025}. ISO-NE historically cleared a fixed quantity before adopting a sloped curve and subsequently evolving toward a probabilistic \emph{Marginal Reliability Impact} (MRI) construct, where the slope is tied to the incremental reduction in EUE \cite{zhao2017constructing}. NYISO uses a multi-segment zonal/system curve with a cap and a zero-crossing quantity, translating between installed capacity (ICAP) and unforced capacity (UCAP) using derate factors \cite{NYISO_ICAP_Demand_Curves}. This heterogeneity is not merely cosmetic: it reflects differing views on how well current curves map to consumer preferences and how closely demand-curve parameters should be benchmarked to the incremental reliability value of capacity.

\subsubsection{The Impact of Data Center Uncertainty}
\label{subsubsec:capacity_uncertainty_mitigation}

AI data-center growth stresses capacity market mechanisms because incremental demand is often concentrated in a small number of large, milestone-dependent projects with uncertain energization dates and realized utilization. As a result, the capacity market demand curve becomes unusually sensitive to load forecast error. High-biased forecasts shift the demand curve outward, raising near-term procurement and clearing prices, while low-biased forecasts suppress prices but increase the likelihood of future scarcity and corrective actions. Geographic concentration further tightens local delivery constraints even when system-wide capacity appears sufficient. Under these conditions, the demand curve functions as the primary \emph{risk-allocation interface}: its slope, cap-and-collar features, and update frequency determine how forecast errors propagate into price volatility, investment timing, and exposure to market power when the system clears near scarcity.

\paragraph{Uncertainty in Demand Curve Estimation and Potential Market Power}
Data-center growth increases not only the level of forecasted peak demand, but also the variance and skewness of forecast error because many projects are discrete, milestone-driven, and geographically clustered. This can make the capacity requirement ``jump'' across auction years, increasing volatility in the clearing price even if the underlying supply stack changes slowly. In tight conditions, the price impact of forecast error is amplified because the market clears near the steep portion of the demand curve, where small quantity shifts translate to large price movements.

When reserve margins tighten, capacity markets can become more susceptible to structural and strategic market power \cite{MonitoringAnalytics2025_PJM_SOM_Q22025}. Under uncertain data center growth, a demand curve calibrated on an aggressive load forecast can push the market toward these tight conditions even before load is realized, increasing the probability that prices are set by scarcity rents or mitigation rules rather than by competitive fundamentals. This is one reason that capacity market design and market power mitigation become more central when load growth is dominated by a small number of uncertain mega-load projects.

\paragraph{Mitigation Levers}
Common design responses aim to reduce the sensitivity of auction outcomes to forecast noise while maintaining investable signals. These include: i)~price caps and collars in the demand curve to limit extreme outcomes when forecasts are wrong or supply is temporarily constrained; ii)~prompt or closer-to-delivery procurement to correct forecast errors with better information and shorter supply response requirements; and iii)~more granular/seasonal products where reliability risk is increasingly seasonal. Operationally, these instruments also reduce the value of strategic behavior by increasing procurement frequency and by narrowing the time window over which a single forecast error can dominate procurement~\cite{iso_ne_capacity_auction_reforms}.

\section{Utility Tariff Design and Cost Allocation}\label{sec:tariff_uncertainty}

Large-scale AI data centers introduce an affordability risk that differs from incremental, diversified load growth: they induce \emph{discrete} and often front-loaded transmission and distribution (T\&D) capacity expansion, while their realized demand trajectory (MW level, ramp profile, and site persistence) remains uncertain. When realized utilization falls below the planning assumption, the associated fixed-cost recovery can be shifted to legacy customers through subsequent rate cases. This section summarizes how U.S.\ jurisdictions are adapting tariff design and cost allocation to i)~reduce speculative build risk, ii)~assign incremental delivery costs to cost-causing entities, and iii)~limit cross-subsidies to residential and small commercial customers.

\subsection{Why ``cost shifting'' arises under cost-of-service regulation}
Under cost-of-service regulation (COSR), retail rates are set by i)~establishing an allowed revenue requirement, ii)~functionalizing and allocating that revenue requirement across customer classes and functional cost ``buckets,'' and iii)~mapping each class responsibility to specific rate elements. A compact representation is
\begin{equation}
RR_y \;=\; O_y \;+\; \delta_y \;+\; r_y\,RB_y \;+\; T_y,
\label{eq:revreq}
\end{equation}
where $RR_y$ is the allowed revenue requirement in year $y$ (\$/year), $O_y$ denotes approved operating expenses (including O\&M and pass-through items treated as operating costs), $\delta_y$ is the year-$y$ book depreciation that increases accumulated depreciation (and therefore reduces rate base), $RB_y$ is the (depreciated original cost) rate base, $r_y$ is the allowed rate of return, and $T_y$ aggregates income/property taxes and other regulatory adjustments (e.g., riders, true-ups).
Rate base evolves according to
\begin{equation}
RB_{y+1} \;=\; RB_y \;+\; I_y \;-\; \delta_y \;-\; \rho_y,
\label{eq:ratebase}
\end{equation}
where $I_y$ denotes new plant additions placed in service and included in rate base, and $\rho_y$ captures net plant retirements/write-offs removed from rate base. Eqs.~\eqref{eq:revreq}--\eqref{eq:ratebase} emphasize the \emph{burnout} property: if new investment slows ($I_y \!\approx\! 0$), then $RB_y$ declines mechanically as depreciation and retirements continue, making sustained utilization important when large assets are built in anticipation of future load~\cite{PLJ_EoI}.

Functionalization is important because customers connected at different voltage levels do not make use of the same set of utility facilities. Residential and small commercial customers are generally allocated costs associated with secondary and primary distribution, as well as appropriate shares of sub-transmission, high-voltage transmission, and generation. Medium commercial and industrial customers typically avoid secondary-distribution costs, while large customers connected at sub-transmission voltage generally avoid both primary and secondary distribution costs. Very large loads connected directly at transmission voltage therefore primarily bear costs associated with high-voltage transmission and, in vertically integrated systems, generation. For hyperscale data centers, the principal cost-pooling concern consequently arises not from lower-voltage distribution facilities that they do not use, but from transmission and generation costs that are traditionally pooled across customers and allocated using load-based determinants such as coincident-peak demand. Creating a distinct hyperscale rate class or special tariff can alter this pooling arrangement by ring-fencing clearly incremental transmission and generation costs and assigning them more directly to the large loads that cause those investments.

Cost shifting arises primarily through the pooling and allocation of these functional cost categories. In practice, costs are organized into buckets $k \in \mathcal{K}$ (e.g., customer-related, demand-related, energy-related, transmission, distribution, and public-purpose costs). Class $c \in \mathcal{C}$ revenue responsibility is then computed via bucket-specific allocators:
\begin{equation}
RR_{c,y} \;=\; \sum_{k\in\mathcal{K}} \alpha_{c,k,y}\, C_{k,y},
\qquad
\sum_{c\in\mathcal{C}} \alpha_{c,k,y} \;=\; 1 \;\; \forall k,
\label{eq:allocation}
\end{equation}
where $C_{k,y}$ is total cost in bucket $k$ during year $y$, and $\alpha_{c,k,y}$ is the corresponding class allocator constructed from cost drivers such as coincident peak demand, non-coincident peak demand, energy usage, number of customers, and voltage level. Traditional COSR typically does not maintain a systematic ``new'' versus ``existing'' separation within a class. As a result, incremental upgrades that enter $RB_y$ can propagate into delivery-related buckets via the return and depreciation components embedded in $C_{k,y}$ and then flow into class responsibilities through Eq.~\eqref{eq:allocation}. When hyperscale-driven upgrades are large relative to the embedded legacy system, these embedded-cost channels can raise delivery charges for customers who did not trigger the upgrades.

This mechanism is not universal: if incremental investments are modest and load growth increases utilization, embedded costs can be diluted and average rates can fall. The hyperscale context elevates upward-pressure risk when i)~capacity is built ahead of realized load (front-loading), ii)~the probability of partial build-out, delayed ramp, or relocation is nontrivial, increasing stranding tail risk, and iii)~the nominal cost of new T\&D and generation equipment has risen well above the embedded cost basis of existing facilities, so marginal additions are recovered at materially higher unit costs.

\subsection{Jurisdictional layers: interconnection, transmission, and retail obligations}

Hyperscale data centers often interconnect at transmission voltages (at or above 69~kV). In ISO/RTO regions, engineering assessment and categorization of network upgrades is frequently performed under federal-jurisdictional transmission processes. However, the end-use customer remains a \emph{retail load}: state commissions determine retail cost allocation, tariff applicability, and the class/rate-element mapping through which transmission and distribution costs are recovered.

Transmission cost recovery applies analogous embedded-cost logic at a regional scale. Annualized transmission revenue requirements are allocated using local, zonal, or regional methods intended to reflect cost-causation principles (costs roughly commensurate with burdens caused or benefits received), yet the incidence varies substantially across regions and methodologies. These allocated transmission charges are then flowed through to retail customers via state-jurisdictional class allocation (Eq.~\eqref{eq:allocation}) and rate design~\cite{glachant2025handbook}.

A related rate-design issue arises when hyperscale data centers provide measurable services or benefits to the bulk power system. Such services should, in principle, be compensated through direct payments or bill credits when their value can be identified and verified. Historically, industrial retail tariffs have more commonly imposed technical performance requirements---for example, acceptable power-factor ranges---than provided explicit compensation for grid-support capabilities. For hyperscale loads, however, their scale and controllability may make explicit valuation of such services increasingly important. A practical approach is to separate wholesale valuation from retail compensation: the ISO/RTO could quantify the value of qualifying grid services and reflect those values in the wholesale grid costs or credits allocated to LSEs; the portion attributable to a specific hyperscale data center could then be unbundled and passed through by the utility, subject to state commission approval, as a corresponding bill credit or other compensation. Such an arrangement preserves the respective wholesale and retail regulatory roles while allowing data centers to receive the economic value of verifiable services they provide to the grid. Compensation should therefore be based on measured service delivery rather than installed capability. Infrastructure installed principally to serve the facility itself should remain the customer's cost responsibility; socialized recovery or long-term contracting is defensible only when an asset demonstrably avoids or substitutes for grid-side investment that otherwise would have been planned.

\subsection{Generation procurement and when ``generation cost shifting'' matters}
The role of generation in cost shifting depends on retail market structure. In retail-competitive jurisdictions (and some large-customer access regimes), hyperscale customers often secure energy via bilateral contracts such as power purchase agreements (PPAs), making generation costs primarily contractual rather than embedded in a regulated utility revenue requirement. In vertically integrated jurisdictions, utility-backed generation procurement undertaken to serve a hyperscale load can enter embedded cost recovery; if the load ramps below forecast or exits early, a stranded-cost dispute can emerge over whether unrecovered balances remain in $RB_y$ and how they are allocated across classes in subsequent proceedings.

Even in competitive procurement settings, other customers are not fully insulated from system-wide wholesale market effects of large demand growth. Accordingly, many regulatory proceedings focus on long-lived network investments (T\&D and network upgrades) that are more directly tied to embedded-cost recovery and class allocation.

\subsection{Regulatory tools to limit cross-subsidies and stranding risk}
Because network upgrades are long-lived assets recovered over multi-decade horizons, commissions have increasingly emphasized mechanisms that reduce the probability that unrecovered balances are reassigned to other classes if a data center ramps slowly, reduces load, or exits within a decade.

\paragraph{Up-front cost responsibility (customer-funded upgrades)}
A first lever is to treat certain facilities as customer-specific rather than system-shared. When service requests exceed standard tariff obligations (e.g., dedicated feeders, non-standard facilities, or site-specific reinforcements), utilities may apply contributions in aid of construction (CIAC), higher deposits, and customer-funded line extensions. In engineering terms, this approach assigns specific capital additions to the triggering customer rather than injecting the full $I_y$ into the pooled rate base used for class allocation.

\paragraph{Long-term commitments (minimum bills, contract demand, take-or-pay)}
Commissions have supported multi-year commitments by very large loads that bind cost recovery to a minimum billing level over a horizon aligned with asset recovery. A generic take-or-pay formulation can be expressed at the billing-period level as
\begin{equation}
\text{Bill}_t \;\ge\; \underline{\text{Bill}}, \qquad \forall t \in \{1,\ldots,H\},
\label{eq:minbill}
\end{equation}
where $\text{Bill}_t$ is the billed amount in billing period $t$ and $\underline{\text{Bill}}$ is a contractual minimum over commitment horizon $H$ (often 10--20 years). This structure reduces exposure to underutilization and lowers the probability that unrecovered fixed costs are reallocated to other classes in future rate cases.

\paragraph{Cost segregation and nonlinear rate design (separate ``buckets'' for hyperscale loads)}
A structural approach isolates incremental costs through special tariffs or contracts that effectively create a parallel COSR track (a separate ``bucket'') for hyperscale loads rather than pooling upgrades into legacy embedded costs. Nonlinear rate design provides a technical reconciliation between marginal-efficiency objectives and distributional constraints. A standard tariff formulation is shown below:
\begin{equation}
\text{Bill}_t \;=\; F \;+\; p_E\,E_t \;+\; p_D\,D_t,
\label{eq:tariff}
\end{equation}
where $F$ is a fixed charge (\$/billing period), $E_t$ is energy (kWh) with unit price $p_E$ (\$/kWh), and $D_t$ is a billing demand determinant (kW, e.g., monthly maximum kW or coincident-peak kW) with unit price $p_D$ (\$/kW). In this framing, $p_D$ (and time-varying demand determinants, when used) can be tuned to reflect capacity-driven marginal costs, while $F$ and/or minimum-bill provisions (Eq.~\eqref{eq:minbill}) recover infra-marginal costs and support class equity without distorting marginal operational incentives.

\subsection{``Behind-the-fence'' strategies and unresolved allocation questions}
Developers increasingly consider behind-the-fence generation (often paired with storage) to reduce interconnection exposure and improve reliability. However, commissions may retain authority over certain overhead-cost allocations and policy obligations (e.g., renewable portfolio standards). Moreover, complete cost isolation is technically and institutionally difficult: network expansions can produce shared benefits (e.g., reduced congestion or improved contingency performance), and stakeholders often dispute what capacity is truly incremental versus what would have been built under baseline planning trajectories. These debates sit at the intersection of cost-causation norms, embedded-cost recovery, and the practical limits of disentangling joint network benefits~\cite{glachant2025handbook}. These questions become sharper if the interconnection itself evolves from a single metered point into a multi-port interface serving generation, storage, and load simultaneously. Cost responsibility is presently assigned at the level of a customer and a service connection, whereas a multi-port architecture would require attribution at the level of individual ports and bidirectional flows---a granularity that the CIAC and take-or-pay constructs above are not designed to express. A regulatory precedent nonetheless exists in distributed energy resources: a rooftop-solar customer is already simultaneously a generator and a load, and rate design accommodates this not through a new customer class but through metering arrangements, i.e., net metering, or separate production and consumption meters. A multi-port interface generalizes this to a multi-meter system in which each port is metered, and potentially charged or credited, according to the direction and nature of its flows; the same architecture therefore governs both what a port is charged for the service it draws and what it may be credited for a service it verifiably delivers. How such a credit could be arranged across the wholesale and retail layers is set out earlier in this section.

The analogy holds at the level of metering and settlement design rather than compensation rates: the net-metering cost-shift debate is itself a caution that such design choices carry significant allocation consequences, and at transmission scale the closer precedents are co-located generation and station-power rules rather than retail distributed energy resource (DER) tariffs. Designing such multi-meter, port-differentiated tariffs is future work; we return to this prospect, and to its dependence on the maturity of the enabling technology, in Section~\ref{sec:framework}.

\subsection{Broader rate design transition and implications}
These issues arise alongside an ongoing rate design transition. Many residential tariffs remain dominated by volumetric charges, whereas large C\&I tariffs more commonly include demand charges and voltage differentiation; performance-based and incentive-oriented mechanisms can further change the timing and incidence of cost recovery~\cite{PLJ_EoI}. Overall, the prevailing direction is not to prohibit growth, but to require large new loads to i)~internalize a larger share of clearly incremental delivery costs, ii)~commit to billing determinants over horizons aligned with asset lives, and iii)~face tariffs with capacity-reflective marginal components, thereby reducing speculative build risk and limiting bill impacts on residential and small commercial customers while preserving efficient marginal price signals.

Sections~\ref{sec:adequacy_uncertainty}--\ref{sec:tariff_uncertainty} emphasized uncertainty at planning and regulatory horizons; we now shift to \emph{variation} at operational horizons, where sub-second dynamics and minute-scale ramps create new security and stability requirements. Section~\ref{sec:security_variation} examines these reliability implications and the mechanisms through which data centers can either exacerbate or mitigate them.

\section{Operational Security and Stability Under Variation}\label{sec:security_variation}
Building on the timescale crosswalk in Section~\ref{subsec:variation}, we focus here on operational reliability under variation: i)~sub-second dynamic performance (power quality, ride-through, oscillations, and stability) and ii)~seconds-to-minutes ramps that drive reserve deployment, congestion management, and operational flexibility requirements. The high power density of AI compute clusters and the prevalence of power-electronic interfaces (UPS systems, rectifiers/inverters, on-site storage and generation) can introduce grid-visible dynamics that are unfamiliar to operators who historically treated load as largely passive.

\subsection{Sub-Second Dynamics: Transients, Power Quality, Ride-Through, and Oscillatory Stability}\label{subsec:transient}

The rapid emergence of large data centers has altered demand profiles and influenced the operations of modern power systems. At sub-second time scales, data centers exhibit complex transient behaviors driven by power-electronic circuits and high-speed control loops. It is therefore essential to investigate these transients in detail and examine their role in shaping power quality, fault ride-through capability, and oscillatory stability.

In Table~\ref{tab:dc-load}, we compare data centers (power-electronics loads) with classical loads and highlight how data centers differ in governing mechanisms, harmonics, nonlinearity, fluctuation, inertia, and modeling characteristics. More details are provided in the subsequent discussions. It is clear that prior models and intuition are no longer adequate to address the challenges of data center integration.

\begin{table}
\caption{Comparison of Classical Loads and Data Centers}
\label{tab:dc-load}
\centering
\setlength\tabcolsep{3pt}
\begin{threeparttable} 
\begin{tabular}{p{1.8cm} >{\raggedright\arraybackslash}p{3.2cm} >{\raggedright\arraybackslash}p{3.2cm}}
\toprule
                    & \textbf{Conventional Loads}          & \textbf{Data Centers}               \\
\midrule
Mechanism   & mostly physics-driven (passive, loose regulation)   & control-driven (active, tight regulation)     \\
Harmonics           & limited                           & rich (covering low to high frequencies)            \\
Nonlinearity        & limited                           & highly nonlinear (waveform distortion, saturation) \\
Fluctuation         & slight to moderate                & huge and rapid                                     \\
Inertia             & low to high                       & almost zero       \\
Characteristics  & constant impedance ($P \propto V^2$, dominant), & constant power ($P$ fixed, negative impedance)   \\
                    & constant current ($P \propto V$),  &                                                    \\
                    & constant power ($P$ fixed)        &                                                    \\
\bottomrule
\end{tabular}
\end{threeparttable}
\end{table}

\subsubsection{Transient Dynamics}

Transient dynamics of data centers refer to the milli- to sub-second variations of voltage/current waveforms or power swings in response to external disturbances~\cite{taylor2012voltage}. As a result, these dynamics tend to trigger sophisticated interactions between data centers and the utility grid. They can damage sensitive electronic facilities (e.g., compute servers, communication networks), trigger protective devices (e.g., fuses, protection relays), or even cause cascading system failures. Transient dynamics are uniquely able to capture real-time operational details of data centers, which are unobservable in other long timescales (steady state). In particular, these details can provide a full picture of nonlinear characteristics and fast-acting responses (e.g., harmonics, current saturation) during faulty conditions. It turns out that a deep understanding of data center transients is the key to mitigating risks and ensure reliable operation.

Specialized models have been developed to capture the transient dynamics. Two common modeling options are root-mean-square (RMS) models and EMT models, and a detailed comparison is provided in Table~\ref{tab:rms-emt}. RMS models apply electric phasors, which are vectors representing a sinusoidal component of the fundamental frequency (60/50 Hz), to analyze sinusoidal dynamics. Dq-frame formulations~\cite{lin2021input} are widely used to transform the sinusoids into algebraic constants. This formulation is effective at capturing slowly varying waveforms under balanced or weakly unbalanced conditions. While computationally efficient, it cannot represent dynamic modes during asymmetrical faults or rapid voltage variations due to the neglect of high-frequency harmonics.
On the other hand, EMT models capture the full process of electromagnetic transients, including saturation, harmonics, converter switching, and fast control loops. For instance, in \cite{Samanta2024}, an EMT model was developed for large cryptocurrency mining loads. Another aggregate model combining resistive loads, switch-mode power supply loads, and lighting loads was formulated in \cite{milani2022detailed}. EMT models are essential for tracking data center transients, but industry-grade EMT models for data centers are presently lacking. Other modeling options include frequency-domain models (which rely on local linearization) and averaging models (which focus on low-frequency dynamics)~\cite{emadi2004modeling}. 

Modern data centers utilize multiple stages of power conversion from utility feeders (AC) to terminal chips (DC)~\cite{chen2022data}. In this regard, their transient behaviors are largely driven by hardware known as power-electronic converters (AC-to-DC rectifiers and DC-to-DC choppers). Formally, data center dynamics are governed by two essential aspects of power electronics: topology and control~\cite{zhang2022model}. There is a clear industry trend toward reducing the number of power conversion stages to improve overall efficiency. Novel topologies along with wide bandgap semiconductors allow for higher switching frequencies and near-peak efficiency~\cite{sandri2017increasing}. Despite rapid advances in hardware innovation, the dynamical performance of these new converters has not been thoroughly explored, and many of them may exhibit different (or troublesome) behaviors in sub-second timescale. Additionally, control designs play an essential role in data center transient dynamics. From a feedback control perspective, the tight regulation of power electronic converters will shape the feedback loops and govern the convergence/evolution of interconnected systems. Many converters in data centers implement ``constant power'' mode, which isolates compute chips from input voltage fluctuations but introduces negative impedance to the upstream grid (thus degrading stability performance)~\cite{emadi2004modeling}. 

\begin{table}
\caption{Comparison of RMS and EMT Models}
\label{tab:rms-emt}
\centering
\setlength\tabcolsep{3pt}
\begin{threeparttable} 
\begin{tabular}{p{1.8cm} >{\raggedright\arraybackslash}p{3.2cm} >{\raggedright\arraybackslash}p{3.2cm}}
\toprule
            & \textbf{RMS Models}   & \textbf{EMT Models}   \\
\midrule
Variables    & voltage/current phasors     & instantaneous voltages/ currents  \\
Modeling     & electromechanical oscillations, voltage/frequency variations  & electromagnetic transients, switching, harmonics \\
Computation & efficient, highly scalable   & intensive, poorly scalable   \\
Accuracy    & moderate precision    & highest precision   \\
Use Cases   & sinusoidal (quasi) steady-state, three-phase balanced systems    & broad, including nonlinear, non-steady-state, unbalanced systems \\
\bottomrule
\end{tabular}
\end{threeparttable}
\end{table}

\subsubsection{Power Quality}

Power quality describes how well the electric power supplies (e.g., voltage/current magnitudes, frequencies, waveforms) stay within a desirable range~\cite{eaton2004price}. It can be regarded as the approximate equilibrium of transient dynamics measured at the point of common coupling. A wide range of standardized metrics is used in current practice, including voltage/frequency deviation (sag/swell depth), total harmonic distortion, and voltage unbalance factor.

Most power-quality issues are attributed to the internal power electronics equipment, iron-saturating devices, and arcing devices. In data centers, the involved nonlinear power conversions and nonlinear loads~\cite{hoevenaars2020data} greatly distort the sinusoidal waveforms and inject rich harmonics into the upstream grid. Well-known symptoms include equipment failure and misoperation, increased energy losses and overheating~\cite{miletic2023development}, often resulting in unplanned downtime and reduced system reliability. Power factor correction (PFC) converters, a type of AC-to-DC circuits, become a popular option to enforce the input current waveform sinusoidal and in phase with the input voltage (close to unity power factor)~\cite{zhu2023adaptive}. Another option is hybrid transformers, which combine a traditional low frequency transformer with a power electronic module for harmonic isolation~\cite{prabowo2020control}. The current practices show that, by employing proper mitigation strategies, data centers can achieve energy savings while maintaining the stability, purity, and cost-efficiency of the power supply.

Power quality modeling must account for the above intricate dynamics while allowing rapid validation of performance across diverse topological and control designs. RMS models are effective for analyzing voltage magnitude and unbalance status but they lack the fidelity to capture harmonics and fast dynamics. There are harmonic-based extensions that replace a single-frequency representation with a weighted sum of harmonic components. EMT models provide the highest level of modeling fidelity and track the full details of switching events and harmonics. Furthermore, the current infrastructure has introduced a new power-quality issue of supraharmonics in the 2--150\,kHz range~\cite{sutaria2023propagation}, typically generated by high-frequency control signals.

\subsubsection{Fault Ride-Through}

Fault ride-through describes the ability to remain connected and operational for a short period (e.g., 150 ms) during voltage or frequency disturbances~\cite{piya2018fault}. These disturbances may come from a nearby fault or an uncontrolled fast transient or surge. As defined, fault ride-through is a resilience measure that mandates operational support during external disturbances~\cite{liu2021active}. Fault ride-through has historically been a requirement for interconnected generators, but recent practices have begun to extend these rules to large loads, including data centers~\cite{ercot2025lel}.

Data centers face unique physical and operational hurdles during external disturbances because of their intense deployment of power electronics. Specifically, most power-electronic equipment acts as a constant-power load. It increases current draw in response to voltage drops (a positive feedback), which can dangerously strain a weak grid and hinder disturbance rejection. If massive loads trip simultaneously, the abrupt power imbalance can induce sharp voltage and frequency spikes, which greatly stress the grid and increase the risk of cascading failures~\cite{conto2025texas}. Physically, power-electronic converters are inherently sensitive to voltage variations, and without specific control, they may disconnect compute loads during voltage sags or swells. Many power electronic circuits do not equip enough energy capacity reserve to enable reactive power injection and prevent the rapid voltage collapse~\cite{arghandeh2012flywheel}. Backup power and storage are found useful in extending the fault ride-through performance~\cite{eaton2004price}. There are challenges around the unpredictable transient behaviors for power electronics as well, and rigorous site-specific tests are often needed to manage the recovery process and grid reconnection dynamics.

The power industry is shifting toward active management of data centers for system supports. It becomes increasingly common for grid codes to require system participants to inject reactive power during faults~\cite{yan2017smart}. Grid operators are also advocating for better site-specific load modeling and transparency \cite{samanta2024prediction}. There is also a growing trend for data centers to evolve into large controllable loads instead of passive consumers.

\subsubsection{Oscillatory Stability}

An electric grid is oscillatory stable if it can maintain synchronization and acceptable operating conditions. To achieve this, it must adequately damp all oscillations in the system triggered by arbitrary disturbances. The integration of data centers brings new stability risks to the grid, primarily due to the wide timescale dynamics and frequency-coupling effects of converters~\cite{wang2018harmonic}. Oscillations can arise from interactions between multiple data centers or between data centers and the electric grid. The negative impedance and weak grid conditions are two major contributors to instability. In recent years, the power swings associated with AI training workloads have raised new concerns regarding high-magnitude, periodic power fluctuations and their impacts on system stability and power quality~\cite{ko2025wide,valverde2025forced}.

Data centers and other converter-based resources have motivated a revisit of the existing power system stability categories. The task force~\cite{hatziargyriou2020definition} has identified two new stability categories that exhibit a close connection to power-electronics loads and generators. The first category, converter-driven stability, covers instabilities driven by the cross-coupling between converters and the grid electromagnetic transients. The periodic, structured power fluctuations driven by repetitive training patterns may serve as a forcing signal to trigger such stability issues~\cite{jimenez2025data}. The second category is resonance stability (harmonic stability), and it focuses on the waveform distortions both above and below the fundamental frequency. In data centers, the harmonic components might be generated and poorly damped at certain frequencies.

Integrated analytical methods are required to evaluate oscillatory stability in the presence of data centers. The common tools include small-signal and eigenvalue analysis, impedance-based methods, and EMT simulations. Small-signal analysis can provide rich information on oscillation modes and participation factors of system parameters. In \cite{sun2022data,sun2022data2}, single-phase power supply units were rigorously modeled to capture DC bus dynamics and the coupled current responses. The new model then supported a more accurate assessment of system stability. A well-known drawback of small-signal analysis is its reliance on linearized dynamics, and this method is often limited to certain operating conditions. Impedance-based methods model different components by their frequency-dependent impedance and investigate the interaction between source and load impedance instead of the full state spaces. While this method is flexible and allows convenient validation through measurements, it still depends on typical operating points and becomes inaccurate during large disturbances. Reference~\cite{ruan2025data} used a neural network to model impedance under a large-signal setting. The data-driven model provided an effective solution to study the adverse interactions between data centers and the grid. EMT simulation is the most precise and time-consuming of all the options. It is particularly suited for power-electronics-rich systems and weak-grid conditions when other methods fail to capture the true dynamics.

A further difficulty is scale. The screening methods above are typically applied to a small number of interacting terminals, whereas a converter-dense facility with many independently controlled ports---some regulating a DC bus, some tracking maximum power, some exchanging reactive power with the grid---constitutes a substantially higher-dimensional coupled system. Whether impedance-based screening, including its data-driven variants, remains both tractable and appropriately conservative at that scale is an open question; hierarchical screening, aggregated port-level impedance models, and model-order reduction are plausible directions, but validated practice does not yet exist. The same limitation applies across the interface as well as within it: neither the facility nor the upstream system can be screened in isolation, and validated interaction models spanning the two do not yet exist. A practical decomposition nonetheless follows from the interface itself: for grid-interaction screening, what must be characterized is the facility's aggregate impedance seen at the point of interconnection, with internal stability certified separately by the facility. This does not dissolve the internal problem, but it assigns each half to the party able to observe it and makes the interconnection-facing quantity something that can be specified and verified.

\subsection{Seconds-to-Minutes Variation: Ramping Products, Reserves, and Operating Flexibility}\label{subsec:ramping}

At operational timescales, a dominant concern is the net-load ramp (and rebound) imposed at the point of interconnection by coordinated workload changes, inference bursts, and potentially correlated behavior across multiple facilities (including geographic failover). These ramps translate directly into the deployment of regulation and ramping reserves, and they can exacerbate congestion in regions where large loads cluster~\cite{sheehan_microgrid_dc_density}. In practice, operators manage this through look-ahead dispatch, explicit ramp procurement, or both.

Advanced power-electronic interfaces are sometimes expected to remove this requirement altogether by absorbing fluctuations before they reach the grid. That expectation should be bounded. The buffering capability of a solid-state interface is limited by its current headroom---on the order of $1.5\times$ rated, against roughly $25\times$ for two seconds in a line-frequency transformer---and by millisecond-scale device thermal constants, so it can reshape the spectral signature seen upstream but cannot absorb the energy-bearing ramps discussed here~\cite{guillod2017protection}. Any masking that does occur additionally creates a measurement problem, because the ramp observed at the meter no longer reveals the stress borne by the interface itself; this argues for qualification and telemetry requirements that observe internal state rather than terminal flow alone. The deeper point is that the two domains are not separable. Dynamics inside the facility and dynamics on the upstream system are coupled through the interface, and the coupling runs in both directions: the interface reshapes what the grid observes, while conditions at the point of interconnection determine how much the interface can absorb at any instant. Characterizing either side alone is therefore insufficient, and transient behavior at the boundary has to be coordinated by design rather than assumed away. This is also why the intrinsic characterization in Section~\ref{subsec:spectral_validation} remains grid-relevant rather than internal: it is the source description that any interaction model across the interface requires.

\paragraph{Reserve-relevant ramps and congestion exposure (seconds--60\,min)}
Coordinated job starts/stops, checkpointing phases, ramp-bounded controls, and post-curtailment rebound can produce ramps that must be balanced by committed ramping capability and operating reserves~\cite{RMI2024}. Data centers can worsen this requirement when ramps are uncoordinated, or partially mitigate it when controllability is explicit through workload orchestration and on-site buffers (UPS/BESS energy shifting, temporary throttling, or limited deferral)~\cite{Alapera2019EEM}. 

To illustrate the operational relevance, consider CAISO's Flexible Ramping Product (FRP), which procures upward and downward ramping capability to cover the forecast net-load change \emph{plus} a probabilistic uncertainty margin over rolling 5- and 15-minute intervals~\cite{CAISOFRP2020,WangHodge2017FRPReview}. During steep solar ramp periods, CAISO's duck-curve analysis identifies a system requirement to supply an additional 13{,}000\,MW within approximately three hours as solar output declines~\cite{CAISO2016DuckCurve}, corresponding to an average sustained ramp of roughly 70\,MW/min across that window. Similarly, ERCOT's Contingency Reserve Service (ECRS) and Responsive Reserve Service (RRS) are sized based on the largest credible contingency~\cite{ERCOT2024ECRS}. Production telemetry from hyperscale training clusters indicates that, in extreme cases, aggregate power consumption can oscillate by tens of megawatts within a single data center, with the amplitude growing in proportion to the number of synchronized GPUs and their peak per-GPU power~\cite{choukse2025power}. Swings of this magnitude are not yet comparable to the largest credible contingency, but they are large enough that the associated reserve and ramping burden ceases to be negligible as facility scale increases. Operational descriptors therefore include MW/min ramp rates, reserve sufficiency margins, scarcity price adders, binding-constraint indicators, and response-time requirements for demand response participation~\cite{ERCOT2023LFLTF}. A key question is whether the procured ``ramp capability'' can actually be delivered as a feasible multi-interval trajectory when the system is tight.

\paragraph{Diurnal/event-driven structure and sustained loading (1--24\,h)}
Across hours, inference load can track human activity and exogenous events, imprinting diurnal and event-driven structure that aligns with system stress periods and amplifies congestion exposure even when annual energy is unchanged~\cite{E32024DataCenters}. Training is more schedulable and can be shifted toward low-price/low-carbon intervals, but rebound after curtailment or deferred execution can re-concentrate demand and reshape net-load and congestion patterns~\cite{Google2020CarbonIntel,Google2023DR,Google24x7}. Sustained high-utilization loading also couples to thermal limits and cooling-system inertia. Unlike traditional industrial loads with diurnal cool-down periods, AI training runs can persist at near-peak power for days or weeks, denying transformers the thermal relief typically assumed in IEEE/IEC loading guides and accelerating insulation degradation~\cite{IEEEC5791,IEEE2019TransformerAging}. The coupling to cooling-system inertia, where thermal storage or chiller scheduling can buffer short-term thermal transients, offers a partial mitigation pathway~\cite{IceCooling}.

\paragraph{Mechanisms for demand response and ancillary services.}
Practical pathways for data centers to provide demand response and ancillary services include workload shifting/deferral, controlled throttling (e.g., GPU frequency scaling), and leveraging UPS systems (and co-located BESS where present) as short-duration buffers to shape ramps and reduce reserve deployment~\cite{AlKez2021,Ratnam2020}. Recent demonstrations show that coordinated UPS discharge and workload modulation can deliver frequency regulation response within seconds, potentially qualifying data centers for fast-response reserve products~\cite{Alapera2019EEM}. Because flexibility can be time- and state-dependent (e.g., limited by job deadlines, thermal headroom, or UPS state of charge), operational programs must explicitly represent response speed, duration, and rebound characteristics rather than treating flexibility as a static MW quantity. These demand-response mechanisms are moving from concept to field validation: DOE's Data Center Load Flexibility workshop convened operators, utilities, and national laboratories to scope near-term demand-response opportunities~\cite{DOE2025DCFlex}, and EPRI's DCFlex initiative is now running multi-year demonstrations of demand response and workload/load shifting at hyperscale facilities across a planned portfolio of roughly ten sites in the U.S.\ and Europe~\cite{EPRI2024DCFlex}.

\paragraph{Market design and qualification: pricing response characteristics with verifiable performance.}
Whether existing reserve products correctly price the response characteristics that matter for data-center flexibility (response time, ramp rate, sustainable duration, and rebound effects) remains an open question. FRP-style products compensate resources for maintaining headroom rather than for energy throughput, aligning well with data center capabilities where the goal is to avoid rather than \emph{deliver} rapid load swings~\cite{CAISOFRP2020}. Recent comparisons of look-ahead dispatch and ramp-product implementations highlight two practical gaps: rolling-window operation can drift into different end states, and ramp awards may not guarantee intertemporal deliverability unless additional constraints are enforced; proposed remedies include enforcement terms (e.g., ramp-increment/rolling-difference constraints) and, in some systems, a mid-duration ramp product to cover extended ramps under stress~\cite{zhang2026comparativeassessmentlookaheadeconomic}. However, qualification requirements (telemetry granularity, baseline methodologies, performance scoring) must ensure that offered flexibility is measurable and verifiable while remaining compatible with data center service-level agreements~\cite{CRS2025}. In practice, qualification rules should align incentives with system needs (e.g., fast response during scarcity hours, congestion-aware delivery) and reduce the risk that nominal flexibility fails to materialize under tight operating conditions~\cite{NERC2024SRA}. As data center loads grow to represent a significant fraction of regional demand, market designs may need to evolve toward dynamic, location- and time-differentiated reserve requirements that explicitly account for the load-side contribution to system ramping needs~\cite{NERC2024ReserveMargin}.

\section{Affordability Under Variation: Grid Reinforcement and New Flexibility Requirements}\label{sec:affordability_variation}
Variation has an affordability dimension because it changes the type and quantity of assets needed to meet reliability standards. A grid built to serve a high-load-factor but slowly varying demand profile may be inadequate or expensive to retrofit when faced with large, fast ramps, voltage-sensitive behavior, and concentrated power-electronic loads. As discussed in Sec.~\ref{subsec:ramping}, these operational choices shape what ``counts'' as usable flexibility; that, in turn, changes which investments are cost-effective (e.g., storage duration, monitoring, and power-conditioning upgrades). This section identifies the incremental infrastructure and operational capabilities that may be required, from substation and transmission reinforcements to flexibility resources such as energy storage and dynamic reactive power. We then connect these engineering requirements to cost assessment and allocation, emphasizing how tariff and contracting choices determine whether the least-cost portfolio is ultimately built.

\subsection{Network and Interconnection Upgrades for High-Density, Fast-Ramping Loads}\label{subsec:network}
AI data centers often arrive as large step changes at specific substations, driving upgrades in transformers, breakers, protection, feeders, and upstream transmission to manage thermal limits and contingency flows. The integration of hyperscale AI facilities introduces a fundamental discontinuity in grid planning due to extreme spatial load density, often exceeding 50~kW per rack for GPU-intensive clusters and reaching approximately 120~kW for a fully populated {NVIDIA} {GB200} {NVL72} rack \cite{schneider_wp110_ai,nvidia_gb200_rack_power}, and the rapid temporal volatility of computational workloads.

These characteristics necessitate significant upgrades to physical infrastructure. As noted in Section~\ref{subsec:ramping}, sustained training can deny transformers the thermal relief that loading guides assume, accelerating insulation ageing~\cite{LBNL2024}. Inference workloads additionally create ``bursty'' demand profiles that can fluctuate by tens of megawatts per minute at individual facilities, inducing rapid thermal cycling and mechanical stress on substation equipment. Consequently, utilities must deploy advanced thermal monitoring and dynamic rating systems to maximize the utilization of existing transformer capacity while avoiding accelerated aging. On the facility side, emerging wide-bandgap power electronics (GaN, SiC) are enabling more efficient rectification and power delivery, potentially reducing upstream infrastructure stress \cite{navitas_gan_sic}.

Because these upgrades are capital-intensive and can have multi-year lead times (currently up to four years to secure large power transformers, roughly double 2021 levels~\cite{iea2025transmissiongrid}), they raise option-value questions: should infrastructure be built ahead of demonstrated need, or incrementally as projects materialize? The traditional ``just-in-time'' investment model risks creating speculative queues and delaying energization, forcing developers toward carbon-intensive bridging solutions such as backup diesel generation. Conversely, anticipatory investment reduces lead times but increases the risk of stranded assets if projected loads fail to materialize. Recent studies suggest that coordinating transmission planning with data center siting can reduce total system investment costs by over 12\% compared to uncoordinated expansion \cite{zhang_2025_joint_planning}.

To mitigate these risks, modular interconnection designs and ``Connect-and-Manage'' strategies are emerging as viable alternatives. Drawing on approaches pioneered in the UK's regulatory framework \cite{ofgem_connect_manage}, similar concepts are being explored in U.S.\ markets: data centers may connect prior to the completion of deep network reinforcements by agreeing to ``non-firm'' access or flexible operational limits. For instance, a facility might be granted early energization with a contractual obligation to curtail load during specific grid contingencies. This approach trades capital investment (new lines) for operational complexity (active curtailment management), requiring robust software orchestration to ensure reliability margins are maintained without compromising the stability of the sub-transmission network \cite{edf_flexible_interconnection}. It also increases the risk of cost shifting to other customers or of stranded assets when cost shifting is constrained by regulators.

\subsection{Flexibility and Power-Quality Investments: Storage, Grid-Forming Controls, and Filtering}\label{subsec:flex_assets}

Mitigating the variability introduced by AI data centers is best framed as a portfolio problem spanning three complementary resource classes: energy storage, demand-side flexibility (including demand response and workload shifting), and grid-forming control. This framing is increasingly reflected in guidance from international agencies and industry consortia. The IEA's \emph{Energy and AI} report identifies flexible data-center operation and co-located storage as central levers for reconciling data-center growth with grid reliability, cost, and decarbonization goals~\cite{IEA2025EnergyAI}. In parallel, EPRI's DCFlex initiative, launched in 2024 and now spanning grid operators (including ERCOT and PJM), utilities, and hyperscalers (Google, Meta, Microsoft, NVIDIA), is running multi-year field demonstrations of storage, demand response, and load-shifting strategies across a planned portfolio of roughly ten flexibility hubs in the U.S.\ and Europe, explicitly aimed at transitioning data-center backup power and on-site flexibility into dependable grid assets~\cite{EPRI2024DCFlex,DOE2025DCFlex}. The remainder of this subsection details the engineering basis for each mitigation pathway.

\subsubsection{Storage and Grid-Forming Control}

Energy storage, and batteries in particular, are key resources for enhancing data center flexibility and power quality~\cite{he2025performance}. They can reshape facility demand profiles and mitigate rapid fluctuations; the appropriate design depends on what the system operator compensates and can verify. Fast, short ramps call primarily for power capability, while longer ramps and rebound management require sufficient energy duration and appropriate controls (Sec.~\ref{subsec:ramping}). Advanced control of on-site batteries can provide inertia-like behavior that supports grid frequency stability and helps maintain voltage stability during disturbances. Storage systems can also mitigate voltage sags, harmonic distortion, and flicker caused by rapidly fluctuating AI training traces. Storage capacity should be selected through a careful trade-off between economic efficiency and the required level of system reliability. Scenario analysis and uncertainty-aware planning can evaluate the benefits of additional storage deployment and compliance with grid requirements for mandatory transient-performance criteria.

Hybrid energy storage is a powerful strategy for combining technologies to exploit complementary performance and cost benefits~\cite{zheng2016hybrid}. Leveraging diverse storage devices (typically supercapacitors and batteries) can provide high power density and high energy density simultaneously. Hierarchical scheduling of these resources can improve performance, extend operational lifetime, and reduce premature failures~\cite{wang2023hierarchical}.

Grid-forming control is a novel mechanism that transforms energy storage from a passive backup into an active, grid-stabilizing asset~\cite{shamseldein2025liability}. Under grid-forming control, batteries act as independent voltage sources and regulate voltage and frequency. They can provide inertia-like response and fault ride-through support to the broader system. For instance, \cite{azizi2025grid} found that parallel grid-forming batteries can help mitigate voltage sags and flicker, alleviating power-quality concerns. On-site UPS batteries have also been modified to implement grid-forming control that supports fault isolation and protection~\cite{qi2022modeling}. Grid-forming batteries are increasingly being deployed and are valued for providing voltage/frequency control while enhancing power quality and grid resilience~\cite{ross2026using}. To ensure that grid-forming storage from different vendors can be qualified and interconnected reliably at scale, the U.S.\ Department of Energy-supported UNIFI Consortium has published vendor-agnostic technical specifications defining system- and inverter-level requirements for grid-forming resources, providing a standards basis for treating grid-forming storage as a dependable mitigation asset for large, fast-ramping loads such as AI data centers~\cite{UNIFI2024GFM}.

\subsubsection{Power Quality and Filtering}
Fast variation can necessitate incremental investments in flexibility and power quality: short-duration storage for ramps and frequency response; dynamic reactive power devices (e.g., static synchronous compensators (STATCOMs) or synchronous condensers) for voltage support; and harmonic filtering and monitoring to manage interactions between power-electronic equipment and the grid.

The interaction between the grid and the nonlinear switch-mode power supplies (SMPS) of thousands of GPU servers creates a complex power-quality (PQ) environment. These loads inject harmonic currents, particularly triplen harmonics (3rd, 9th) and characteristic 5th/7th orders. These injections can cause neutral conductor overheating and voltage distortion at the Point of Common Coupling (PCC)~\cite{eaton_harmonics}. Unlike passive loads, AI clusters require active mitigation strategies.
\begin{itemize}
    \item Active Harmonic Filtering (AHF): Due to the dynamic nature of AI workloads, static passive filters are often insufficient and may create resonance. Active filters, while more capital intensive, dynamically inject counter-currents to cancel harmonic distortion across a wide spectrum, helping ensure compliance with the IEEE 519 standard.
    \item Dynamic Reactive Power Support: The rapid ramp rates of AI clusters can cause severe voltage sags. Traditional capacitor banks are too slow to respond ($>1$ second), and their output drops with the square of voltage ($Q \propto V^2$). Advanced STATCOMs can provide millisecond-level voltage support and, when paired with grid-forming controls, deliver virtual inertia to stabilize frequency during large load steps \cite{hitachi_statcom}.
\end{itemize}

A key boundary decision is whether mitigation is provided behind the meter (on-site storage and power conditioning) or on the system side (utility-installed devices). Distributed, behind-the-meter (BTM) mitigation adheres strictly to cost-causation principles but may result in capital inefficiency due to over-provisioning of redundancy at each site. Conversely, centralized utility-side mitigation benefits from economies of scale but raises regulatory challenges regarding cost socialization.

The comparison above shows how cost allocation and tariff structures can either encourage efficient mitigation or unintentionally socialize avoidable costs. Modern ``Large Load Tariffs'' are increasingly incorporating strict power factor penalties, minimum demand charges (e.g., 85--90\% take-or-pay clauses), and CIAC fees to ring-fence these costs \cite{lbl_rate_design_2025}. Furthermore, innovative tariffs that incentivize flexibility can transform these loads from liabilities into grid assets, reducing the need for expensive peaking generation. However, discrete engineering upgrades and tariff adjustments alone are insufficient; aligning these physical and financial layers necessitates the integrated techno-economic approach explored in Section~\ref{sec:framework}.

\section{Toward a Techno-Economic Policy Framework}\label{sec:framework}

\subsection{A Broader Structural Change is Coming}

Although this paper is focused on engineering and economic aspects, it is important to emphasize that the future trajectory of the AI--power grid nexus remains highly uncertain. This uncertainty reflects not only technical factors, such as rapid evolution in AI architectures and efficiency; but also broader geopolitical dynamics, industrial policy, and the pace of progress in off-grid or hybrid generation resources. We hypothesize that regions with stronger, more resilient, and more adaptable power grid capabilities are more likely to attract sustained, long-term investment in AI data centers. Relatedly, regions that are able to co-design digital infrastructure with inter-temporal flexibility aligned to local grid characteristics are likely to be better positioned in the emerging global competition for AI capacity.

From an education and workforce development perspective, these dynamics point to a broader structural shift. We hypothesize that AI infrastructure will emerge as a major domain of convergence for electrical engineering, computer science, and economics. This convergence creates an opportunity to develop high-impact curricula that go beyond a narrow focus on AI algorithms or data analysis; instead, we must train students to reason across the full infrastructure stack, spanning power systems, computing architecture, markets, and policy. This underpins the development of advanced AI and, potentially, artificial general intelligence.

\subsection{Potential Solution: PowerAgent-X}

Large AI data centers provide a timely and consequential use case for illustrating the value of an agentic, systems-level perspective on the AI--power grid nexus. They are among the fastest-growing sources of electricity demand, with significant implications for grid capacity, reliability, emissions, and governance. At the same time, they possess atypical operational flexibility that makes them natural candidates for coordinated, grid-aware operation \cite{xie2026crucial}. In regions such as PJM, where large-load interconnection has become a focal reliability and policy challenge, the binding constraint is often not the absence of engineering options, but the lack of trusted, scenario-based analysis capable of evaluating complex, human-in-the-loop pathways under deep uncertainty, including contested forecasts of future data center demand and the system impacts of gigawatt-scale load additions.

Building on recent advances in agentic AI and the PowerAgent community \cite{zhang2025poweragent}, we propose \emph{PowerAgent-X}: an agentic simulation framework designed to study grid planning and interconnection decision-making under data-center-driven uncertainty. The key departure from conventional planning models is that coordination failures arise endogenously. Instead of assuming a single planner with perfect foresight, PowerAgent-X represents utilities, system operators, regulators, communities, and data-center developers as heterogeneous agents who learn over time, operate under institutional procedures, and may delay, disagree, or act strategically. This matters for data-center growth because uncertainty is \emph{milestone-driven} and geographically clustered: small delays in queue processing, transmission upgrades, or demand coordination can redirect interconnection outcomes, shift congestion patterns, and trigger costly stopgap measures.

\begin{figure}
\centering
  \includegraphics[width=0.4\textwidth]{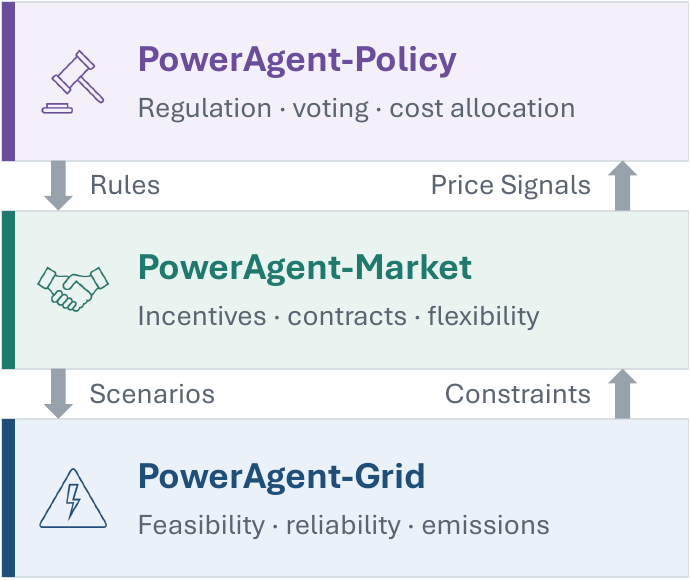}
  \caption{The PowerAgent-X framework and the roles of its three coupled layers. PowerAgent-Grid evaluates physical feasibility, reliability, and emissions; PowerAgent-Market represents participant behavior, contracting, and flexibility decisions; and PowerAgent-Policy represents governance, cost allocation, and siting processes. Bidirectional interfaces allow physical constraints to shape incentives and policy while market and policy decisions alter the scenarios evaluated by the grid layer.}
  \label{poweragentx}
\end{figure}

As shown in Fig.~\ref{poweragentx}, PowerAgent-X couples three layers through explicit interfaces so that physical constraints shape incentives, incentives shape behavior, and governance determines whether coordination occurs in time. The bottom layer, \emph{PowerAgent-Grid}, provides an engineering-faithful feasibility backbone that translates candidate interconnection locations, timelines, and operating profiles into congestion, deliverability constraints, reliability risk, and emissions outcomes using trusted grid tools in an agentic workflow. The middle layer, \emph{PowerAgent-Market}, models how participants respond to rules and incentives through queue entry strategies, contracting, and flexibility commitments; it also represents AI-aware data-center control (e.g., workload shifting and temporal load modulation) that reacts to prices, reliability signals, and carbon intensity. The top layer, \emph{PowerAgent-Policy}, simulates regulatory and stakeholder processes that frequently dominate timelines, including negotiations over cost allocation, siting, reliability requirements, and community acceptance under asymmetric information.

A practical advantage of this architecture is that it supports mechanism verification for planning under uncertainty. For example, planners can evaluate how alternative interconnection rules, flexibility products, or cost-allocation policies change i)~queue outcomes and upgrade timing, ii)~the credibility and timing of data-center flexibility, and iii)~downstream reliability and emissions impacts. The agentic market layer is particularly useful because it makes high-dimensional preferences and constraints elicitable and simulatable at scale: by converting concise natural-language descriptions of risk tolerance, operational constraints, and contractual priorities into structured experimental economics model \cite{wang2025experimental}, the framework can generate auditable distributions of bids, commitments, and strategic responses that are difficult to obtain from purely optimization-based models.

PowerAgent-X therefore provides a unified environment for diagnosing and mitigating the planning challenges posed by data centers: it can quantify the reliability and cost consequences of delayed transmission and interconnection, identify incentive designs that elicit verifiable flexibility, and surface policy bottlenecks that determine whether technically feasible solutions are implemented on time. This enables transparent comparisons between uncoordinated growth and coordinated, AI-aware pathways in which regulatory enablement and performance-based commitments reduce congestion, accelerate deployment, and lower long-run system costs.

{

\subsection{Potential Solution: Advanced Grid Integration}

New developments are emerging in the grid integration technologies for future AI data centers. An exciting trend is that the interface between the grid and data centers will evolve from a single connection point into a sub-network.

\begin{figure}
    \centering
    \includegraphics[width=\linewidth]{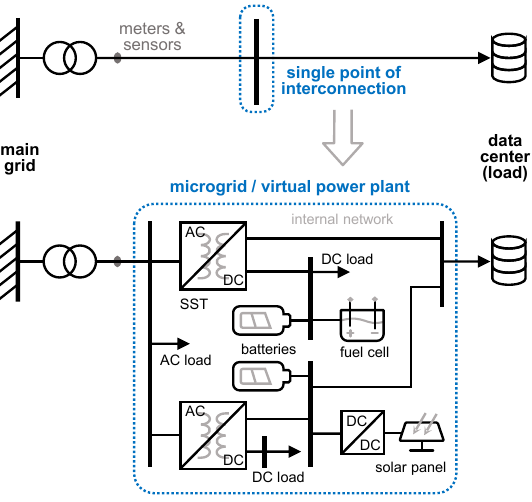}
    \caption{Conventional and advanced grid--data-center interfaces. The upper diagram shows a data center connected to the grid through a single metered point of interconnection. The lower diagram expands that boundary into an internal microgrid/virtual-power-plant sub-network in which solid-state-transformer-based multi-port converters can coordinate AC and DC loads, batteries, a fuel cell, and solar generation. The architecture permits bidirectional flows among the internal ports while retaining meters and sensors at the external grid boundary.}
    \label{fig:interface}
\end{figure}

Fig.~\ref{fig:interface} shows an illustrative layout for the new kind of interfaces. In the future grid integration, there will be an interface expansion to form local microgrids~\cite{li2016toward} or virtual power plants (VPPs)~\cite{ruan2024data} with internal networks and a number of distributed energy resources. Bidirectional and fluctuating energy flows will alter the coupling between the grid and data centers, and ultimately change whole-system performance.

Fig.~\ref{fig:interface} also highlights the enabling technology, namely solid-state transformers (SSTs)~\cite{huber2026next}. By design, SSTs integrate power electronics with high-frequency transformers, which can offer greater operational flexibility, multi-port integration of heterogeneous resources, and---in applications where they also displace rectification stages, as in DC-distribution data centers---improved end-to-end conversion efficiency and a more compact footprint. Ongoing advances in SSTs are therefore closely tied to the feasibility of the microgrid/VPP interface described above.

These prospects must, however, be weighed against the present cost and maturity of the technology, which bound how quickly such interfaces can be expected at scale. On a per-kVA basis SSTs remain considerably more expensive than line-frequency transformers (LFTs): matched 7.2~kV medium-voltage prototypes place the SST at $86.2$~\$/kVA against $25$~\$/kVA for an LFT, a factor of $3.4$~\cite{zheng2022cost}, while assessments of AC--AC distribution duty in the 1000~kVA class report a premium of at least $5\times$ together with roughly $3\times$ higher losses~\cite{huber2019applicability}. A component-level comparison is nonetheless incomplete, because an SST also displaces rectification, filtering, and reactive-power compensation equipment; on a functional-equivalence basis, and particularly in DC-dominant data centers, the effective premium is materially smaller than the headline multiple suggests. The non-cost barriers are arguably more binding. An LFT can deliver on the order of $25\times$ rated current for two seconds, whereas an SST is typically limited to about $1.5\times$ with millisecond-scale thermal constants at the die, so overcurrent-based protection coordination does not carry over unchanged~\cite{guillod2017protection}. Existing test codes cannot presently assign a basic insulation level to a medium-voltage SST, which by itself precludes drop-in substitution within an insulation-coordinated distribution system~\cite{xu2022insulation}, and standards for medium-frequency insulation and partial-discharge measurement remain under development~\cite{ozdemir2021insulation}. Deployment reflects these constraints: SSTs are not in routine service on utility distribution feeders, with field experience concentrated in railway traction, a small number of medium-voltage demonstrations, and pilot fast-charging installations~\cite{saleh2019partii,byrne2024reliability}. We therefore treat SST-based interfaces as a \emph{conditional long-term option} rather than a near-term substitute for conventional interconnection: Fig.~\ref{fig:interface} indicates a direction of travel, while the timing depends on cost trajectories, protection practice, and standardization. This framing matters for the regulatory questions raised below, since port-level cost attribution and premium products for grid-forming service become pressing only as, and to the extent that, the underlying hardware is actually deployed.

In addition, the new interfaces require an upgrade of the energy management process and toolboxes. There should be a new layer of microgrid/VPP control between the existing grid-side and data center management systems. Note that this extra layer is occasionally viewed as an energy community in the literature.
The new energy management process and toolboxes will continue to emerge and represent a direction for future development. This trend is driven by the higher penetration of distributed energy resources and, conditionally, by the anticipated maturation of SST technology. There will be significant challenges and opportunities in the future, including complex operating status, flexibility resources, market design, policy making, and practical implementation.

\subsubsection{Complex Operating Status and Flexibility}

The increased number of power flow paths and system components has created a strong need for coordinated operation. As introduced above, a new coordination layer (centralized or distributed) must integrate advanced controllers/schedulers to enhance flexibility resources and mitigate operational risks of power fluctuations and harmonics. Through these coordination layers, many data centers can shift their role from passive loads to proactive grid assets.

Flexibility potential can be unlocked through SSTs~\cite{huber2026next} because they allow full control of power flow directions and multi-port operations, and they are able to smooth, and partially whiten, the fluctuations of data center demand seen upstream---though only within the current and thermal limits noted in Section~\ref{subsec:ramping}, so that the underlying variability is reshaped rather than eliminated. Flexibility can also come from other technologies, e.g., grid-forming batteries~\cite{ross2026using}. These facilities provide frequency and voltage regulation services for enhancing the whole-system stability margins.

\subsubsection{Market and Policy Making}

Unlike the single-point interconnection, the internal sub-network is expected to operate under complex status, which increases the burden for measurement and allocation of energy costs. This issue becomes even more pronounced within a distributed decision-making framework. There is a clear desire to redesign the settlement mechanisms to efficiently coordinate between the external grid operators and the internal energy consumers.

A two-layer settlement structure organizes this. In the upper layer the facility settles with the utility as a single entity, and prevailing practice retains a single metered point of interconnection even when the internal network is extensive, precisely because measurement and billing are tractable there; the cost-responsibility instruments of Section~\ref{sec:tariff_uncertainty} continue to apply unchanged at that boundary. What the multi-port architecture changes is the lower layer, where energy-flow tracking, loss allocation, and port-level attribution arise, and for which bilateral contracting and network-usage-based allocation are the established candidates.

A notable trend is the emergence of new market operators and business models. The new operators may serve as a local balancing agent or a load aggregator. They have strong incentives to innovate their business models to serve different energy stakeholders, e.g., community-based markets and peer-to-peer trading~\cite{sousa2019peer}.

Meanwhile, policy makers need new analysis and design tools to regulate these new markets and business practices. Typical issues include how to incentivize demand response participation, how to ensure fair revenue allocation among stakeholders, and how to leverage subsidies to encourage investment in new flexibility and grid-supporting technologies.

\subsubsection{Towards Practical Implementation}

Sufficient support for practical implementation is of great importance. For instance, a major concern associated with the new grid integration pathway is around the potential risks and uncertain benefits. In this respect, PowerAgent-X may provide a forward-looking assessment of these new issues by adapting its grid-market-policy modeling capacity to local microgrids/VPPs. In addition, industrial standardization and workforce training need to align with future trends of technological development and engineering practices.

Ultimately, the energy community must maintain an open mindset and actively explore diverse technologies and business models to navigate a future of both challenges and opportunities. This mindset will help us identify the optimal pathway that well serves AI and meets energy sustainability goals. There is little doubt that technological breakthroughs alone will not determine future success. Instead, a holistic approach that integrates new technologies, new business models, and new policy interventions will become essential for shaping the pathway forward.

}

\section{Conclusion} 
\label{sec:conclusion}

Large data centers are emerging as a major and qualitatively distinct source of electricity demand. Their integration into power systems challenges conventional planning, operational, and regulatory frameworks because large facilities arrive as spatially concentrated point loads with uncertain realization and exhibit workload-driven variability across multiple time scales. This paper organized these challenges around two defining load properties---\emph{forecast uncertainty} and \emph{inter-temporal variation}---and examined how they propagate into reliability and affordability outcomes.

Using a four-quadrant framework linking reliability and cost to uncertainty and variation, we showed that AI data center growth affects power systems across planning and operational horizons. At longer time scales, uncertainty in size, timing, location, and firmness complicates resource adequacy assessment, generation and transmission expansion, and cost allocation under cost-of-service regulation and capacity market designs. At shorter time scales, coordinated and power-electronic-driven variability introduces new requirements for operational security, stability, and ramping reserves. These effects are amplified by the spatial concentration of data centers, which limits diversification and increases the local consequences of forecast and operational errors.

A key insight from this analysis is that engineering, economic, and institutional dimensions are tightly coupled. Reliability risks are shaped not only by physical constraints but also by tariff structures, interconnection rules, and contracting mechanisms that determine how uncertainty is revealed and how commitments are enforced. Similarly, affordability outcomes depend on technical choices regarding mitigation location, flexibility deployment, and the coordination of generation, transmission, and load-side investments. Treating these dimensions in isolation can lead to inefficient infrastructure build-out, misaligned incentives, or unintended cost shifts.

The paper also highlighted that AI data centers are not inherently inflexible loads. Empirical characterization of workload-driven power variability suggests that, particularly for training-dominated facilities, a substantial share of variation occurs at time scales compatible with demand response, reserve provision, and congestion management. Realizing this potential, however, requires explicit modeling, measurement, and verification of response characteristics---including speed, duration, rebound behavior, and deliverability---rather than relying on simplified notions of interruptibility.

Several research directions follow from this work. First, resource adequacy and planning models would benefit from improved representations of load-side uncertainty, including milestone-dependent realization, correlated siting decisions, and conditional flexibility availability. Second, further development and validation of dynamic models for large power-electronic loads are needed to assess stability and interaction risks as data center penetration increases. Third, market and tariff designs should be evaluated with explicit attention to how forecast error and spatial concentration affect procurement outcomes, investment incentives, and cost allocation. Finally, integrated simulation frameworks that combine physical constraints with institutional behavior offer a promising approach for studying coordination challenges under deep uncertainty.

More broadly, the growth of AI data centers illustrates a shift toward electricity demand that is increasingly shaped by digital systems, software-driven control, and rapid technological change. Addressing this shift will require analytical approaches that integrate power system engineering, economics, and policy, and that can adapt as both computational workloads and power systems continue to evolve.

\section*{Disclaimer}
The views expressed in this paper are those of the authors and do not reflect the views of PJM Interconnection, L.L.C. or its Board of Managers, of which Le Xie is a member.

\bibliographystyle{IEEEtran}
\bibliography{ref}
\end{document}